\documentclass{article}
\usepackage{epsfig}
\catcode`\@=11
\def\slash{\mathpalette\make@slash}
\def\make@slash#1#2{\setbox\z@\hbox{$#1#2$}%
  \hbox to 0pt{\hss$#1/$\hss\kern-\wd0}\box0}
\catcode`\@=12 

\begin{document}

\title{
 \vskip-3cm{\baselineskip14pt
  \centerline{\normalsize\hfill TTP99--43}
  \centerline{\normalsize\hfill hep-ph/9910332}
 }
 \vskip.7cm
 {\Large\bf{ Renormalization and Running of Quark Mass and Field 
 in the Regularization Invariant and $\overline{\mathrm{MS}}$ Schemes   
 at Three and Four Loops}
 \vspace{1.5cm}
 }
}

\author{{K.G. Chetyrkin}\thanks{Permanent address:
Institute for Nuclear Research, Russian Academy of Sciences,
60th October Anniversary Prospect 7a, Moscow 117312, Russia.}
\  and 
A. R\'etey
  \\[3em]
  {\it Institut f\"ur Theoretische Teilchenphysik,} \\
  {\it Universit\"at Karlsruhe, D-76128 Karlsruhe, Germany}
}

\date{}
\maketitle

\begin{abstract}
We derive explicit transformation formulae relating the renormalized
quark mass and field as defined in the $ \overline{\mathrm{MS}}
$-scheme with the corresponding quantities defined in any other
scheme. By analytically computing the three-loop quark propagator in
the high-energy limit (that is keeping only massless terms and terms
of first order in the quark mass) we find the NNNLO conversion factors
transforming the $ \overline{\mathrm{MS}} $ quark mass and the
renormalized quark field to those defined in a ``Regularization
Invariant'' ($ \mathrm{RI} $) scheme which is more suitable for
lattice QCD calculations.  The NNNLO contribution in the mass
conversion factor turns out to be large and comparable to the previous
NNLO contribution at a scale of 2 GeV --- the typical normalization
scale employed in lattice simulations. Thus, in order to get a precise
prediction for the $ \overline{\mathrm{MS}} $ masses of the light
quarks from lattice calculations the latter should use a somewhat
higher scale of around, say, 3 GeV where the (apparent) convergence of
the perturbative series for the mass conversion factor is better.

We also compute two more terms in the high-energy expansion of the $
\overline{\mathrm{MS}} $ renormalized quark propagator. The result is
then used to discuss the uncertainty caused by the use of the high
energy limit in determining the $ \overline{\mathrm{MS}} $ mass of the
charmed quark.  As a by-product of our calculations we determine the
four-loop anomalous dimensions of the quark mass and field in the
Regularization Invariant scheme.  Finally, we discuss some physical
reasons lying behind the striking absence of $\zeta(4)$ in these
computed anomalous dimensions.
\end{abstract} 

\section{Introduction}\label{sec:intro}

Quark masses are fundamental parameters of the QCD Lagrangian.
Nevertheless, their relation to measurable physical quantities is not
direct: the masses depend on the renormalization scheme and, within a
given one, on the renormalization scale $\mu$.

In the realm of pQCD the definition which is most often used is based
on the $ \overline{\mathrm{MS}} $-scheme \cite{ms,MSbar} which leads
to the so-called short-distance $ \overline{\mathrm{MS}} $ mass.  Such
a definition is of great convenience for dealing with mass-dependent
inclusive physical observables dominated by short distances (for a
review see \cite{CKKrep}).  Unfortunately it is usually difficult to
get precise information about the quark masses from predictions from
these considerations, as their mass dependence is relatively weak.

To determine the absolute values of quark masses, one necessarily has
to rely on the methods which incorporate the features of
nonperturbative QCD. So far, the only two methods which are based on
QCD from first principles are QCD sum rules and Lattice QCD (for
recent discussions see e.g.
\cite{Prades:1997vn,Jamin:1997sa,Dominguez:1998zs,Narison:1999mv,Chetyrkin:1998ej,Prades:1998xa,Lubicz:1998kc,Kenway:1998ew,Gimenez:1998pu,Sharpe:1998hh}.
Rather accurate determinations of the ratios of various quarks masses
can be obtained within Chiral Perturbation Theory \cite{leut1}.

Lattice QCD provides a direct way to determine quark masses from first
principles.  Unlike QCD sum rules it does not require model
assumptions.  It is possible to carry out a systematic improvement of
Lattice QCD so that all the discretization errors proportional to the
lattice spacing are eliminated (a comprehensible review is given in
\cite{Luscher:1998pe}). The resulting quark mass is the (short
distance) bare lattice quark mass. The matching of the lattice quark
masses to those defined in a continuum perturbative scheme requires
the calculation of the corresponding multiplicative renormalization
constants.  In the $ \mathrm{RI} $ scheme \cite{NMPmethod} the
renormalization conditions are applied to amputated Green functions in
Landau gauge, setting them equal to their tree-level values.  This
allows the non-perturbative calculation of the renormalization
constants. An alternative to the $ \mathrm{RI} $ approach is the
Schroedinger functional scheme (SF) which was used in
\cite{Capitani:1997mw,Sint:1998iq}.

An impressive number of various lattice determinations of quark masses
recently has been performed (see
Refs~\cite{mq_lanl,mq_fnal,mq_ape96,mq_sesam,mq_apetov,mq_qcdsf,mq_ape98,mq_noi,mq_alpha,qcdsf,mq_jlqcd,mq_cppacs,mq_dwall}).

Once the $ \mathrm{RI} $ quark masses are determined from lattice
calculations they can be related to the $ \overline{\mathrm{MS}} $
mass by a corresponding conversion factor. By necessity this factor
can be defined and, hence, computed only {\em perturbatively\/}. The
conversion factor is presently known at next-to-next-to-leading order
(NNLO) from \cite{Franco:1998bm}. The NNLO contribution happens to be
numerically significant. This makes mandatory to know the NNNLO
$O(\alpha_s^3)$ term in the conversion factor.

In the present article we describe the calculation of this term. It
turns out that the size of the newly computed term is comparable to
the previous one at a renormalization scale of 2 GeV --- the typical
scale currently used in lattice calculations of the light quark
masses. This means that perturbation theory can not be used for a
precise conversion of the presently available $ \mathrm{RI} $ quark
masses to the $ \overline{\mathrm{MS}} $ ones. A simple analysis shows
that the convergence gets much better if the scale is increased to,
say, 3 GeV.  Thus, once the lattice calculations produce the $
\mathrm{RI} $ quark masses at this scale our formulas will allow an
accurate conversion to the $ \overline{\mathrm{MS}} $ masses at the
same scale.

The article is organized as follows.  In section \ref{sec:massfield}
we discuss the scheme dependence of the quark field and mass and
present a general procedure to find corresponding conversion factors
from one scheme to another. The general technique is illustrated by
constructing the conversion factors between the $
\overline{\mathrm{MS}} $ and the $ \mathrm{RI} $ scheme.  In section
\ref{sec:quarkprop3l}, we present the first few terms of the small
mass expansion of the three-loop quark-propagator in the $
\overline{\mathrm{MS}} $ scheme. In section \ref{sec:RIvsMSandMOM} we
first present the conversion functions for the quark mass and field,
and then investigate the validity of the massless approximation for
these functions. Then we use these results to calculate the anomalous
dimensions for the quark mass and field in the $ \mathrm{RI} $ scheme,
the so-called RG invariant mass $\hat{m}$.  The final section is
devoted to conclusions.

In Appendices A and B we display our results for the small mass
expansion of the fermion propagator and the various conversion factors
with their full dependence on the group theoretical factors $C_F$,
$C_A$ and $T $. In Appendix C the four loop anomalous dimensions of
the quark mass and field are listed for the case of a $ \mathrm{SU}
(N)$ gauge group.

Our main results are also available as ASCII input for the programs
{\tt FORM} and {\tt Mathematica} at the following internet address:\\
{\tt http://www-ttp.physik.uni-karlsruhe.de/Progdata/ttp99/ttp99-43/ }

\section{Scheme dependence of quark mass and field}
\label{sec:massfield}

\subsection{Generalities}
\label{sub:general}

We start by considering the bare quark propagator (for simplicity we
stick to the Landau gauge in this section and, thus, do not explicitly
display the gauge dependence)
\begin{equation}
\label{def:fermi}
S_0(q,\alpha_s^0,m_0)
=
\mathrm{i} \int \mathrm{d}x \, \mathrm{e}^{\mathrm{i}qx}
\langle \mathrm{T}[\psi_0 (x) \bar\psi_0 (0)] \rangle
 =  \frac{1}{m_0  - \slash{q} -  \Sigma _0}
{},
\end{equation}
with the quark mass operator $ \Sigma _0$ being conveniently
decomposed into Lorentz invariant structures according to
\begin{equation}
\label{def:sigmas}
 \Sigma _0 = \slash{q}  \Sigma ^0_V + m_0  \Sigma ^0_S  \nonumber  {},
\end{equation} 
where $m_0$ and $\psi_0$ are the bare quark mass and field,
respectively.  Additionally we are using the common shortcuts for the
coupling constant throughout this article:
\[
a^0_s
\equiv  \frac{\alpha^0_s}{\pi}
= \frac{g_0^2}{4\pi^2}
\, ,
\]
$g_0$ being the bare QCD gauge coupling. To be precise we assume that
(\ref{def:fermi}) is dimensionally regulated by going to non-integer
values of the space-time dimension $D=4-2 \epsilon $
\cite{dim.reg-a,dim.reg-b}. The $ \overline{\mathrm{MS}} $
renormalized counterpart of the Green function (\ref{def:fermi}) reads
\begin{eqnarray}
 \displaystyle 
S(q,\alpha_s,m,\mu) 
& = &
\mathrm{i} \int \mathrm{d}x \, \mathrm{e}^{\mathrm{i}qx}
\langle \mathrm{T}[\psi (x) \bar\psi (0)] \rangle
=  \frac{1}{m  - \slash{q} -  \Sigma } 
\nonumber \\
& = &
 \displaystyle  Z_2^{-1} S_0(q,\alpha_s^0,m_0)|_{{}_{
\scriptstyle
m_0=Z_m m,\, \alpha_s^0= \mu^ \epsilon  Z_\alpha \alpha_s } } {},
\label{def:fermi-ren}
\end{eqnarray}
where the renormalized quark field is
\[
\psi= Z_2^{-1/2} \psi_0
{}
\]
and the {}'t Hooft mass parameter $\mu$ is the scale at which the
renormalized quark mass is defined.  The renormalization constants
$Z_2,Z_\alpha$ and $Z_m$ are series of the generic form

\begin{equation}
Z_? = 1 + \sum_{i>0} Z_?^{(i)} \frac{1}{ \epsilon ^i}\, \, \, \, \, \, , \, \,
Z_?^{(i)} = \sum_{ j \ge i } Z_?^{(i,j)} \left(  \frac{\alpha_s}{\pi}  \right )^j {}.
\label{def:Z-expansions}
\end{equation}

A general theorem (first rigorously proven for minimal subtraction in
\cite{Breitenlohner:1977hr,Breitenlohner:1977te}) states that there is
a unique choice of renormalization constants of the form
(\ref{def:Z-expansions}) which makes the propagator\footnote{In fact
all Green functions if proper renormalization constants for gluon and
ghost fields are introduced.}  finite in the limit of $D \to 4$.

The independence of the bare coupling constant, mass and quark field
on $\mu$ leads in the standard way to the following renormalization
group equations for their renormalized counterparts
\begin{eqnarray}
\mu^2\frac{\mathrm{d}}{\mathrm{d}\mu^2} 
\frac{\alpha_s(\mu)}{\pi}
|_{{\displaystyle g_{\scriptscriptstyle {0}},
                           m_{\scriptscriptstyle {0}}
}   }
& = & \beta(\alpha_s) \equiv
-\sum_{i\geq 0}\beta_{{i}}\left(\frac{\alpha_s}{\pi}\right)^{i+2}
= \sum_{i > 0} (i) Z_\alpha^{(1,i)} 
{},
\label{def:beta}
\end{eqnarray}
\begin{eqnarray}
\mu^2\frac{\mathrm{d}}{\mathrm{d}\mu^2} 
 {m}(\mu)|{{}_{\displaystyle g_{0},
 m_{0} }}
& = &
 m(\mu)\gamma_m(\alpha_s) \equiv -
 {m}\sum_{i\geq0}\gamma_m^{(i)}
 {}
 \left(\frac{\alpha_s}{\pi}\right)^{i+1} \nonumber \\
& = & 
m(\mu)  a_s  \frac{\partial Z_m^{(1)}}{\partial  a_s }
\label{def:anomal-mass}
\end{eqnarray}
and
\begin{eqnarray}
 2\mu^2\frac{\mathrm{d}}{\mathrm{d}\mu^2} \psi(\mu)|{{}_{\displaystyle 
 \psi_0, g_{0}, m_{0} }}
& = & 
 \psi(\mu)\gamma_2(\alpha_s) \equiv - \psi(\mu)\sum_{i\geq0} 
 \gamma_2^{(i)} {} \left(\frac{\alpha_s}{\pi}\right)^{i+1}
\nonumber \\
& = &
 \psi(\mu)  a_s  \frac{\partial Z_2^{(1)}}{\partial  a_s }{}.
\label{def:anomal-wave}
\end{eqnarray}

Now let us consider the quark propagator renormalized according to a
different subtraction procedure. Marking with a prime parameters of
the second scheme one can write
\begin{eqnarray}
 \displaystyle 
S'(q,\alpha'_s,m',\mu) 
& = & 
\mathrm{i} \int \mathrm{d}x \, \mathrm{e}^{\mathrm{i}qx} 
\langle \mathrm{T}[\psi' (x) \bar\psi' (0)] \rangle 
 = \frac{1}{m'  - \slash{q} -  \Sigma '}
\nonumber \\
& = &
 (Z'_2)^{-1} S_0(q,\alpha_s^0,m_0)|_{ {}_{\scriptstyle
 m_0=Z'_m m, \,  \alpha_s^0= \mu^ \epsilon  Z_\alpha' \alpha_s} }
 \label{def:fermi-ren-prime} {},
\end{eqnarray}
where without essential loss of generality we have set $\mu' = \mu$.
The finiteness of the renormalized fields and parameters in both
schemes implies that, within the framework of perturbation theory, the
relation between them can uniquely be described as follows
\begin{equation}
m =  C_m \cdot m' \label{def:Cm} 
\end{equation}
\begin{equation}
\psi = \sqrt{C_2} \cdot \psi'  \label{def:C2} {},
\end{equation}
with the ``conversion functions'' being themselves {\em finite\/}
series in $\alpha_s'$, i.e.
\begin{equation}
C_?  \equiv        
1 + \sum_{i>0} C_?^{(i)} \left( \frac{\alpha_s'}{\pi}\right)^i 
{}
\label{def:C-expansions-ap}
\end{equation}
for $?=m$ or $2$.

Note that in general the coefficients $C_?^i$ may depend on the ratio
$m'/\mu$.  If such a dependence is absent then the corresponding
subtraction scheme is referred to as a ``mass independent'' one.  In
what follows we mainly limit ourselves to considering this latter
case. In addition, being only interested in the conversion functions
$C_2$ and $C_m$, we will assume that the function $C_\alpha$ has
already been determined and, thus, will deal with the following
representation of $C_2$ and $C_m$ in terms of the $
\overline{\mathrm{MS}} $ coupling constant $\alpha_s$:
\begin{equation}
C_?  \equiv        
1 + \sum_{i>0} C_?^{(i)} \left(  \frac{\alpha_s}{\pi} \right)^i 
{}.
\label{def:C-expansions-a}
\end{equation}

The running of $m'$ and $\psi'$ is governed by the corresponding
anomalous dimensions $\gamma_m( a_s )$ and $\gamma_2( a_s )$. A direct
use of Eqs.~(\ref{def:Cm},\ref{def:C2}) gives
\begin{equation}
\gamma'_m = { \gamma_m - \beta \frac{\partial}{\partial  a_s } \ln C_m }
\label{form:gamma-pm} {},
\end{equation}
\begin{equation}
\gamma'_2 = { \gamma_2 - \beta\frac{\partial}{\partial a_s } \ln C_2 }
\label{form:gamma-p2} {}.
\end{equation}
At last, from Eq.~(\ref{def:fermi}) it is easy to see that 
\begin{equation}
S(q) = C_2 \cdot  S'(q) = \frac{C_2}{ m' ( 1 -  \Sigma '_S) - 
\slash{q} (1 +  \Sigma '_V) } 
\end{equation}
or, equivalently,
\begin{equation}
C_2 \cdot ( 1 +  \Sigma _V ) = 1 +  \Sigma '_V \label{form:C2}
\end{equation}
\begin{equation}
C_2 \cdot C_m \cdot ( 1 -  \Sigma _S ) = 1 -  \Sigma '_S . \label{form:Cm}
\end{equation}

The renormalization conditions for the non-$ \overline{\mathrm{MS}} $
scheme should then be used to provide the necessary information about
the right hand side to calculate the conversion factors $C_m$ and
$C_2$ once the $ \overline{\mathrm{MS}} $ renormalized $ \Sigma _V$
and $ \Sigma _S$ are known.

\subsection{Regularization Invariant scheme versus $ \overline{\mathrm{MS}} $}
\label{sub:RIvsMS}

The $ \overline{\mathrm{MS}} $ subtraction scheme is intimately
connected to dimensional regularization and, thus, can not be directly
used with other regularizations, including the lattice one. In
addition, the physical meaning of its normalization parameter $\mu$ is
not transparent and leads to the well-known ambiguities when
considering the decoupling of heavy particles.

It is well-known that the above shortcomings are absent for a wide
class of so-called momentum subtraction ($ \mathrm{MOM} $)
schemes\footnote{ In a sense the oldest subtraction scheme --- the
on-shell one for QED --- can also be considered as an example of a $
\mathrm{MOM} $ scheme.}.  The $ \mathrm{MOM} $ schemes require the
values of properly chosen Green functions with predefined $\mu$
dependent configurations of external momenta to be fixed (usually to
their tree values) independently on the considered order. Practical
calculations can then be performed with any regulator (or even without
it in the regulator-free approach of
\cite{Lowenstein:1974wu,Lowenstein:1974uk}).  A shortcoming of $
\mathrm{MOM} $ schemes is that they are in general not
mass-independent which leads to a complicated running of coupling
constant(s) and mass(es).

A general analysis of the problem of constructing mass-independent
subtraction schemes was performed long ago in
\cite{Weinberg:1973ss}. Following essentially Weinberg's ideas, a
specific example of a mass independent $ \mathrm{MOM} $ scheme for QCD
has recently been considered in \cite{NMPmethod} under the name of $
\mathrm{RI} $ (``Regularization Invariant'') scheme (see also
\cite{Gockeler:1998ye}).  The corresponding renormalization conditions
for $ \Sigma ^{ \mathrm{RI} }_S$ and $ \Sigma ^{ \mathrm{RI} }_V$ read
\begin{equation}
\begin{array}{l}
\lim \limits_{m\rightarrow 0} \frac{1}{48} Tr \left[ \gamma_\mu
    \frac{\displaystyle \partial\left( \slash{q}
    (1+\Sigma^{ \mathrm{RI} }_V)\right)}{\displaystyle \partial q_\mu}
\right]_{q^2=-\mu^2}= 1
\\
\lim \limits_{m\rightarrow 0} \frac{1}{12} Tr \left[ 1 -  \Sigma^{ \mathrm{RI} }_S \right]_{q^2=-\mu^2}
= 1 {},
\\
\end{array}
\label{def:RI-rencond}
\end{equation} 
where the trace is to be taken over Dirac, Lorentz and colour indices.
Note that the zero mass limit in (\ref{def:RI-rencond}) means that
both $ \Sigma ^{ \mathrm{RI} }_V$ and $ \Sigma ^{ \mathrm{RI} }_S$ are
effectively {\em massless\/} functions only depending on the QCD
coupling constant, the normalization point $\mu$ and $q^2$. This also
implies that it is sufficient to compute the $ \overline{\mathrm{MS}}
$ functions $ \Sigma _V$ and $ \Sigma _S$ in massless QCD when
computing the conversion factors from relations (\ref{form:C2}) and
(\ref{form:Cm}).

Application of the renormalization conditions (\ref{def:RI-rencond})
to the conversion formulae (\ref{form:C2}) and (\ref{form:Cm}) leads
to equations that can simply be solved for $C^{ \mathrm{RI} }_m$ and
$C^{ \mathrm{RI} }_2$. All the dependence of $ \Sigma _V$ on $q^2$ is
of the form of $\ell = log(-\frac{q^2}{\mu^2})$, which simplifies the
trace and derivative w.r.t. $q_\mu$ and leads to
\begin{equation}
C^{ \mathrm{RI} }_2 = \left[ 
1 +  \Sigma _V + \frac{1}{2}\cdot \frac{\partial \Sigma _V(\ell)}{\partial \ell} 
\right]^{-1}_{q^2=-\mu^2,\, m=0} 
\label{form:C2RI}
\end{equation}
\begin{equation}
C^{ \mathrm{RI} }_m = \left[ 
\frac{ 1 +  \Sigma _V + \frac{1}{2} \cdot \frac{\displaystyle \partial
 \Sigma _V(\ell)}{\displaystyle \partial \ell}}
     { 1 -  \Sigma _S } \right]_{q^2=-\mu^2\, m=0} {}.
\label{form:CmRI}
\end{equation}
Another quark field renormalization that is useful in numerical
lattice simulations is defined by the condition
\begin{equation}
\lim \limits_{m\rightarrow 0} 
\frac{1}{12} Tr \left[ 1+\Sigma_V^{ \mathrm{RI'} } \right]_{q^2=-\mu^2}= 1 \, {},
\label{def:RIp-rencond}
\end{equation}
which results in the even simpler conversion factors
\begin{equation}
C^{ \mathrm{RI'} }_2 = \left[ 1 +  \Sigma _V \right]^{-1}_{q^2=-\mu^2, \, m=0} 
\label{form:C2RIp}
\end{equation}
\begin{equation}
C^{ \mathrm{RI'} }_m = \left[
		\frac{ 1 +  \Sigma _V }{ 1 -  \Sigma _S }
	     \right]_{q^2=-\mu^2, \, m=0} \, {}.
\label{form:CmRIp}
\end{equation}

Using these equations all conversion factors can easily be obtained,
once the $ \overline{\mathrm{MS}} $ renormalized expressions $ \Sigma
_S$ and $ \Sigma _V$ are known.

It should be noted that in practical lattice calculations the massless
limit on the left hand side of (\ref{def:RI-rencond}) and
(\ref{def:RIp-rencond}) is implemented by choosing $\mu \gg m$. On the
other hand, $\mu$ should be much less than the inverse lattice spacing
$1/a$.  Typically $\mu$ is taken around $2$ GeV.  This means that
lattice determinations do not lead directly to the $ \mathrm{RI} $
quark mass but rather to the mass in a different, mass-dependent,
scheme.  The difference between both schemes can be numerically
non-negligible in the case of the charmed quark.

Thus, for both the $ \mathrm{RI} $ and the $ \mathrm{RI'} $ scheme it
is suggestive to introduce their mass-dependent counterparts $
\mathrm{MOM} $ and $ \mathrm{MOM'} $ as defined by the same
Eqs.~(\ref{def:RI-rencond}) and (\ref{def:RIp-rencond}) but without
taking the $m \to 0$ limit.  The corresponding conversion factors to
the $ \overline{\mathrm{MS}} $ scheme read
\begin{equation}
C^{ \mathrm{MOM} }_2 = \left[ 
       1 +  \Sigma _V + \frac{\displaystyle 1}{\displaystyle 2}\cdot q^2 
       \frac{\partial \Sigma _V}{\partial q^2} \right]^{-1}_{q^2=-\mu^2}  \\
\end{equation}
\begin{equation}
C^{ \mathrm{MOM} }_m = \left[ \frac{ 
       1 +  \Sigma _V + \frac{1}{2} \cdot q^2 
       \frac{\displaystyle \partial \Sigma _V}{\displaystyle \partial q^2}}
       { 1 -  \Sigma _S } \right]_{q^2=-\mu^2} 
\end{equation}
and
\begin{equation}
C^{ \mathrm{MOM'} }_2 = \left[ 1 +  \Sigma _V \right]^{-1}_{q^2=-\mu^2}  \\
\end{equation}
\begin{equation}
C^{ \mathrm{MOM'} }_m = \left[ \frac{ 1 +  \Sigma _V }{ 1 -  \Sigma _S } 
\right]_{q^2=-\mu^2}{}\, \, .
\end{equation}
respectively.

\section{Three-loop quark propagator in $\overline{\mbox{MS}}$-scheme}
\label{sec:quarkprop3l}
 
To find the conversion factors for $ \mathrm{MOM} $ and $
\mathrm{MOM'} $ one needs to compute the functions $ \Sigma _V$ and $
\Sigma _S$ including their full mass dependence. A full analytical
result at two-loop level has been obtained only recently in
\cite{Fleischer:1998dw}. An extension of this calculation up to
three-loops is out of reach of present calculational
technologies. Fortunately for the $ \mathrm{RI} $ and $ \mathrm{RI'} $
schemes one effectively only needs to compute {\em massless\/}
three-loop diagrams -- a problem which in principle was solved long
ago in \cite{me81b}. A promising approach to recover the full mass
dependence of the quark propagator seems to be to employ an expansion
in $(m^2/q^2)$ \cite{Kuehn:Barcelona}. Indeed, as has been
demonstrated in
\cite{Chetyrkin:1994ex,Chetyrkin:1997qi,Harlander:1997xa,Harlander:1997kw}
small mass expansions can be a very effective tool for accurate
predictions of mass dependences provided one is not too close to the
threshold ($q^2=m^2$ in our case). The exact two-loop result for the
quark propagator can provide some insight into the accuracy of such an
expansion.

We have analytically computed three terms in the small mass expansion
of the quark propagator to order $\alpha_s^3$.  The calculation has
been done with intensive use of computer algebra programs.  In
particular, we have used {\tt QGRAF}~\cite{qgraf} for the generation
of diagrams and {\tt LMP} \cite{rhdiss} for the diagrammatic small
mass expansion.  The small mass expansion results in products of
massless propagators and massive tadpoles. These have been evaluated
with the help of the {\tt FORM} packages {\tt MATAD}~\cite{Stediss}
and {\tt MINCER}~\cite{mincer2} (A detailed description of the status
of these algebraic programs can be found in~\cite{HarSte:review} ).
It is convenient to write the functions $ \Sigma _{S/V}$ in the
following way:
\begin{equation}
\label{def:sigma-expansion}
 \Sigma _{S/V} = 1 
+ \sum_{i\ge 1}  \Sigma _{S/V}^{(i)} \, \,  a_s ^i(\mu)
\, {}.
\end{equation} 
Our results for the separate contributions in the Landau gauge and in
the $ \overline{\mathrm{MS}} $ scheme read, where we here only keep
the $n_f$ dependence, $n_f$ being the number of light quark flavours,
with one quark flavour of mass $m$ and $n_f-1$ massless quark
flavours\footnote{the expressions including the full dependence on the
gauge parameter and the group theoretical factors $C_A$, $C_F$ and $T$
are given in Appendix \ref{app:quarkprop3l}. Note that the result for
the three loop massless quark propagator can also be found in the
source code of the updated version of the {\tt FORM} program {\tt
MINCER} \cite{JosVer}.}  ($ l_{qm} \equiv \ln(-q^2/\bar{m}^2)$, $
l_{q\mu} \equiv \ln(-q^2/\mu^2)$, $z=m^2/q^2$, and in Landau gauge $
\Sigma _V^{(1)} \equiv 0$):

\begin{eqnarray}\Sigma_V^{(2)} = & & 
\left[
\frac{359}{144} 
-\frac{7}{48}  \,n_f 
-\frac{3}{4}  \,\zeta_{3}
-\frac{67}{48} l_{q \mu}
+\frac{1}{12}  \,n_f l_{q \mu}
\right]
\nonumber\\
 &{+}& z
\left[
\frac{79}{24} 
-\frac{1}{6}  \,n_f 
-\frac{9}{4}  \,\zeta_{3}
- l_{q \mu}
-\frac{1}{2} l_{q m}
\right]
\nonumber\\
 &{+}& z^{2}
\left[
-\frac{331}{216} 
-\frac{1}{24}  \,n_f 
+\frac{10}{3}  \,\zeta_{3}
-\frac{209}{72} l_{q \mu}
-\frac{1}{6}  \,n_f l_{q \mu}
-\frac{19}{12} l_{q \mu}^{2}
-\frac{209}{144} l_{q m}
  \right. \nonumber \\ &{}& \left.  
\phantom{+z^{2}}
-\frac{1}{12}  \,n_f l_{q m}
-\frac{19}{12} l_{q \mu}l_{q m}
-\frac{19}{48} l_{q m}^{2}
\right]
\nonumber\\
 &{+}& z^{3}
\left[
\frac{123}{32} 
+\frac{1}{18}  \,n_f 
+\frac{109}{54} l_{q \mu}
+\frac{7}{36} l_{q \mu}^{2}
+\frac{109}{108} l_{q m}
  \right. \nonumber \\ &{}& \left.  
\phantom{+z^{3}}
+\frac{7}{36} l_{q \mu}l_{q m}
+\frac{7}{144} l_{q m}^{2}
\right]
\nonumber\\
 &{+}& z^{4}
\left[
\frac{668909}{172800} 
+\frac{1}{48}  \,n_f 
+\frac{5093}{2160} l_{q \mu}
-\frac{317}{72} l_{q \mu}^{2}
+\frac{5093}{4320} l_{q m}
  \right. \nonumber \\ &{}& \left.  
\phantom{+z^{4}}
-\frac{317}{72} l_{q \mu}l_{q m}
-\frac{317}{288} l_{q m}^{2}
\right]
\nonumber\\
 &{+}& z^{5}
\left[
\frac{23155073}{2592000} 
+\frac{1}{90}  \,n_f 
+\frac{249353}{10800} l_{q \mu}
-\frac{25553}{1080} l_{q \mu}^{2}
+\frac{249353}{21600} l_{q m}
  \right. \nonumber \\ &{}& \left.  
\phantom{+z^{5}}
-\frac{25553}{1080} l_{q \mu}l_{q m}
-\frac{25553}{4320} l_{q m}^{2}
\right]
{},
\label{SigmaV2M}
\end{eqnarray}

\begin{eqnarray}\Sigma_V^{(3)} = & & 
\left[
\frac{439543}{10368} 
-\frac{12361}{2592}  \,n_f 
+\frac{785}{7776}  \, n_f^{2}
-\frac{8009}{384}  \,\zeta_{3}
+\frac{55}{72}  \,n_f  \,\zeta_{3}
-\frac{79}{256}  \,\zeta_{4}
  \right. \nonumber \\ &{}& \left.  
\phantom{+}
+\frac{1165}{192}  \,\zeta_{5}
-\frac{52321}{2304} l_{q \mu}
+\frac{559}{216}  \,n_f l_{q \mu}
-\frac{13}{216}  \, n_f^{2}l_{q \mu}
+\frac{607}{128}  \,\zeta_{3}l_{q \mu}
  \right. \nonumber \\ &{}& \left.  
\phantom{+}
-\frac{1}{4}  \,n_f  \,\zeta_{3}l_{q \mu}
+\frac{737}{192} l_{q \mu}^{2}
-\frac{133}{288}  \,n_f l_{q \mu}^{2}
+\frac{1}{72}  \, n_f^{2}l_{q \mu}^{2}
\right]
\nonumber\\
 &{+}& z
\left[
\frac{5461}{96} 
-\frac{667}{144}  \,n_f 
+\frac{5}{54}  \, n_f^{2}
-\frac{10025}{288}  \,\zeta_{3}
+\frac{65}{36}  \,n_f  \,\zeta_{3}
+\frac{3}{2}  \,\zeta_{4}
  \right. \nonumber \\ &{}& \left.  
\phantom{+z}
-\frac{6235}{576}  \,\zeta_{5}
-\frac{857}{24} l_{q \mu}
+\frac{185}{72}  \,n_f l_{q \mu}
-\frac{1}{18}  \, n_f^{2}l_{q \mu}
+\frac{123}{8}  \,\zeta_{3}l_{q \mu}
  \right. \nonumber \\ &{}& \left.  
\phantom{+z}
-\frac{3}{4}  \,n_f  \,\zeta_{3}l_{q \mu}
+\frac{9}{16} l_{q \mu}^{2}
-\frac{481}{96} l_{q m}
+\frac{1}{9}  \,n_f l_{q m}
-\frac{3}{4}  \,\zeta_{3}l_{q m}
  \right. \nonumber \\ &{}& \left.  
\phantom{+z}
-\frac{51}{16} l_{q \mu}l_{q m}
+\frac{1}{6}  \,n_f l_{q \mu}l_{q m}
-\frac{111}{64} l_{q m}^{2}
+\frac{1}{12}  \,n_f l_{q m}^{2}
\right]
\nonumber\\
 &{+}& z^{2}
\left[
-\frac{389057}{82944} 
-\frac{397}{1296}  \,n_f 
+\frac{11}{216}  \, n_f^{2}
+\frac{196475}{3456}  \,\zeta_{3}
-\frac{719}{432}  \,n_f  \,\zeta_{3}
  \right. \nonumber \\ &{}& \left.  
\phantom{+z^{2}}
+\frac{35}{36}  \,\zeta_{4}
-\frac{8815}{432}  \,\zeta_{5}
+\frac{19}{216}  \,B_{4}
-\frac{8303}{384} l_{q \mu}
-\frac{1697}{648}  \,n_f l_{q \mu}
  \right. \nonumber \\ &{}& \left.  
\phantom{+z^{2}}
+\frac{17}{216}  \, n_f^{2}l_{q \mu}
-\frac{5177}{144}  \,\zeta_{3}l_{q \mu}
+\frac{47}{18}  \,n_f  \,\zeta_{3}l_{q \mu}
-\frac{4739}{288} l_{q \mu}^{2}
  \right. \nonumber \\ &{}& \left.  
\phantom{+z^{2}}
+\frac{553}{432}  \,n_f l_{q \mu}^{2}
-\frac{1}{18}  \, n_f^{2}l_{q \mu}^{2}
+\frac{973}{288} l_{q \mu}^{3}
-\frac{23}{54}  \,n_f l_{q \mu}^{3}
-\frac{115007}{6912} l_{q m}
  \right. \nonumber \\ &{}& \left.  
\phantom{+z^{2}}
-\frac{3029}{2592}  \,n_f l_{q m}
+\frac{5}{108}  \, n_f^{2}l_{q m}
-\frac{617}{288}  \,\zeta_{3}l_{q m}
+\frac{3}{4}  \,n_f  \,\zeta_{3}l_{q m}
  \right. \nonumber \\ &{}& \left.  
\phantom{+z^{2}}
-\frac{4127}{144} l_{q \mu}l_{q m}
+\frac{35}{36}  \,n_f l_{q \mu}l_{q m}
-\frac{1}{36}  \, n_f^{2}l_{q \mu}l_{q m}
-\frac{157}{64} l_{q \mu}^{2}l_{q m}
  \right. \nonumber \\ &{}& \left.  
\phantom{+z^{2}}
-\frac{3}{8}  \,n_f l_{q \mu}^{2}l_{q m}
-\frac{3923}{384} l_{q m}^{2}
+\frac{287}{1728}  \,n_f l_{q m}^{2}
-\frac{1915}{384} l_{q \mu}l_{q m}^{2}
  \right. \nonumber \\ &{}& \left.  
\phantom{+z^{2}}
-\frac{1}{18}  \,n_f l_{q \mu}l_{q m}^{2}
-\frac{3359}{2304} l_{q m}^{3}
+\frac{11}{864}  \,n_f l_{q m}^{3}
\right]
{},
\label{SigmaV3M}
\end{eqnarray}

\begin{eqnarray}\Sigma_S^{(1)} = & & 
\left[
-\frac{4}{3} 
+ l_{q \mu}
- z
-2 l_{q \mu}z
- l_{q m}z
+\frac{1}{2} z^{2}
+\frac{1}{6} z^{3}
+\frac{1}{12} z^{4}
+\frac{1}{20} z^{5}
\right]
{},
\label{SigmaS1M}
\end{eqnarray}

\begin{eqnarray}\Sigma_S^{(2)} = & & 
\left[
-\frac{5009}{288} 
+\frac{13}{18}  \,n_f 
+\frac{47}{12}  \,\zeta_{3}
+\frac{509}{48} l_{q \mu}
-\frac{4}{9}  \,n_f l_{q \mu}
  \right. \nonumber \\ &{}& \left.  
\phantom{+}
-\frac{15}{8} l_{q \mu}^{2}
+\frac{1}{12}  \,n_f l_{q \mu}^{2}
\right]
\nonumber\\
 &{+}& z
\left[
-\frac{791}{144} 
+\frac{5}{18}  \,n_f 
-\frac{43}{12}  \,\zeta_{3}
-\frac{319}{24} l_{q \mu}
+\frac{7}{18}  \,n_f l_{q \mu}
+\frac{7}{2} l_{q \mu}^{2}
-\frac{1}{3}  \,n_f l_{q \mu}^{2}
  \right. \nonumber \\ &{}& \left.  
\phantom{+z}
-\frac{409}{48} l_{q m}
+\frac{5}{18}  \,n_f l_{q m}
-\frac{9}{4} l_{q \mu}l_{q m}
-\frac{1}{6}  \,n_f l_{q \mu}l_{q m}
-2 l_{q m}^{2}
\right]
\nonumber\\
 &{+}& z^{2}
\left[
\frac{13}{12} 
-\frac{7}{72}  \,n_f 
-\frac{95}{24} l_{q \mu}
+\frac{1}{4}  \,n_f l_{q \mu}
-\frac{49}{12} l_{q \mu}^{2}
-\frac{1}{24} l_{q m}
+\frac{1}{12}  \,n_f l_{q m}
  \right. \nonumber \\ &{}& \left.  
\phantom{+z^{2}}
-\frac{49}{12} l_{q \mu}l_{q m}
-\frac{49}{48} l_{q m}^{2}
\right]
\nonumber\\
 &{+}& z^{3}
\left[
-\frac{451}{324} 
-\frac{5}{54}  \,n_f 
+\frac{829}{216} l_{q \mu}
+\frac{1}{12}  \,n_f l_{q \mu}
+\frac{527}{108} l_{q \mu}^{2}
+\frac{295}{108} l_{q m}
  \right. \nonumber \\ &{}& \left.  
\phantom{+z^{3}}
+\frac{1}{36}  \,n_f l_{q m}
+\frac{527}{108} l_{q \mu}l_{q m}
+\frac{527}{432} l_{q m}^{2}
\right]
\nonumber\\
 &{+}& z^{4}
\left[
-\frac{211333}{20736} 
-\frac{53}{864}  \,n_f 
-\frac{4831}{432} l_{q \mu}
+\frac{1}{24}  \,n_f l_{q \mu}
+\frac{4843}{216} l_{q \mu}^{2}
  \right. \nonumber \\ &{}& \left.  
\phantom{+z^{4}}
-\frac{551}{108} l_{q m}
+\frac{1}{72}  \,n_f l_{q m}
+\frac{4843}{216} l_{q \mu}l_{q m}
+\frac{4843}{864} l_{q m}^{2}
\right]
\nonumber\\
 &{+}& z^{5}
\left[
-\frac{24966073}{864000} 
-\frac{77}{1800}  \,n_f 
-\frac{349147}{3600} l_{q \mu}
+\frac{1}{40}  \,n_f l_{q \mu}
+\frac{3539}{40} l_{q \mu}^{2}
  \right. \nonumber \\ &{}& \left.  
\phantom{+z^{5}}
-\frac{21667}{450} l_{q m}
+\frac{1}{120}  \,n_f l_{q m}
+\frac{3539}{40} l_{q \mu}l_{q m}
+\frac{3539}{160} l_{q m}^{2}
\right]
{},
\label{SigmaS2M}
\end{eqnarray}

\begin{eqnarray}\Sigma_S^{(3)} = & & 
\left[
-\frac{1612847}{6912} 
+\frac{79621}{3888}  \,n_f 
-\frac{191}{729}  \, n_f^{2}
+\frac{150265}{1728}  \,\zeta_{3}
-\frac{755}{216}  \,n_f  \,\zeta_{3}
  \right. \nonumber \\ &{}& \left.  
\phantom{+}
-\frac{1}{54}  \, n_f^{2} \,\zeta_{3}
+\frac{79}{256}  \,\zeta_{4}
+\frac{5}{12}  \,n_f  \,\zeta_{4}
-\frac{6455}{576}  \,\zeta_{5}
+\frac{40329}{256} l_{q \mu}
  \right. \nonumber \\ &{}& \left.  
\phantom{+}
-\frac{6083}{432}  \,n_f l_{q \mu}
+\frac{73}{324}  \, n_f^{2}l_{q \mu}
-\frac{10013}{384}  \,\zeta_{3}l_{q \mu}
+\frac{17}{36}  \,n_f  \,\zeta_{3}l_{q \mu}
  \right. \nonumber \\ &{}& \left.  
\phantom{+}
-\frac{2589}{64} l_{q \mu}^{2}
+\frac{119}{32}  \,n_f l_{q \mu}^{2}
-\frac{2}{27}  \, n_f^{2}l_{q \mu}^{2}
+\frac{65}{16} l_{q \mu}^{3}
-\frac{7}{18}  \,n_f l_{q \mu}^{3}
  \right. \nonumber \\ &{}& \left.  
\phantom{+}
+\frac{1}{108}  \, n_f^{2}l_{q \mu}^{3}
\right]
\nonumber\\
 &{+}& z
\left[
-\frac{303803}{6912} 
+\frac{2969}{648}  \,n_f 
-\frac{25}{324}  \, n_f^{2}
-\frac{178745}{3456}  \,\zeta_{3}
+\frac{313}{54}  \,n_f  \,\zeta_{3}
  \right. \nonumber \\ &{}& \left.  
\phantom{+z}
-\frac{1}{9}  \, n_f^{2} \,\zeta_{3}
-\frac{13}{3}  \,\zeta_{4}
-\frac{9635}{1728}  \,\zeta_{5}
+\frac{2}{9}  \,B_{4}
-\frac{172685}{1152} l_{q \mu}
  \right. \nonumber \\ &{}& \left.  
\phantom{+z}
+\frac{4703}{432}  \,n_f l_{q \mu}
-\frac{5}{81}  \, n_f^{2}l_{q \mu}
+\frac{7493}{192}  \,\zeta_{3}l_{q \mu}
+\frac{17}{36}  \,n_f  \,\zeta_{3}l_{q \mu}
+\frac{367}{6} l_{q \mu}^{2}
  \right. \nonumber \\ &{}& \left.  
\phantom{+z}
-\frac{193}{24}  \,n_f l_{q \mu}^{2}
+\frac{17}{108}  \, n_f^{2}l_{q \mu}^{2}
-\frac{135}{8} l_{q \mu}^{3}
+\frac{14}{9}  \,n_f l_{q \mu}^{3}
-\frac{1}{18}  \, n_f^{2}l_{q \mu}^{3}
  \right. \nonumber \\ &{}& \left.  
\phantom{+z}
-\frac{73547}{768} l_{q m}
+\frac{839}{108}  \,n_f l_{q m}
-\frac{25}{324}  \, n_f^{2}l_{q m}
+\frac{1645}{384}  \,\zeta_{3}l_{q m}
  \right. \nonumber \\ &{}& \left.  
\phantom{+z}
+\frac{5}{6}  \,n_f  \,\zeta_{3}l_{q m}
-\frac{431}{32} l_{q \mu}l_{q m}
-\frac{391}{144}  \,n_f l_{q \mu}l_{q m}
+\frac{5}{54}  \, n_f^{2}l_{q \mu}l_{q m}
  \right. \nonumber \\ &{}& \left.  
\phantom{+z}
-\frac{167}{16} l_{q \mu}^{2}l_{q m}
+\frac{1}{3}  \,n_f l_{q \mu}^{2}l_{q m}
-\frac{1}{36}  \, n_f^{2}l_{q \mu}^{2}l_{q m}
-\frac{2359}{96} l_{q m}^{2}
  \right. \nonumber \\ &{}& \left.  
\phantom{+z}
+\frac{17}{18}  \,n_f l_{q m}^{2}
-10 l_{q \mu}l_{q m}^{2}
-\frac{9}{2} l_{q m}^{3}
+\frac{1}{9}  \,n_f l_{q m}^{3}
\right]
\nonumber\\
 &{+}& z^{2}
\left[
\frac{882619}{82944} 
-\frac{3151}{1728}  \,n_f 
-\frac{35}{648}  \, n_f^{2}
-\frac{34153}{1728}  \,\zeta_{3}
+\frac{91}{144}  \,n_f  \,\zeta_{3}
-3  \,\zeta_{4}
  \right. \nonumber \\ &{}& \left.  
\phantom{+z^{2}}
-\frac{55}{54}  \,\zeta_{5}
-\frac{131681}{6912} l_{q \mu}
+\frac{3293}{432}  \,n_f l_{q \mu}
-\frac{13}{72}  \, n_f^{2}l_{q \mu}
-\frac{3251}{576}  \,\zeta_{3}l_{q \mu}
  \right. \nonumber \\ &{}& \left.  
\phantom{+z^{2}}
-\frac{1203}{32} l_{q \mu}^{2}
+\frac{35}{432}  \,n_f l_{q \mu}^{2}
+\frac{5}{72}  \, n_f^{2}l_{q \mu}^{2}
+\frac{5401}{432} l_{q \mu}^{3}
-\frac{49}{54}  \,n_f l_{q \mu}^{3}
  \right. \nonumber \\ &{}& \left.  
\phantom{+z^{2}}
+\frac{47095}{13824} l_{q m}
+\frac{2285}{864}  \,n_f l_{q m}
-\frac{2}{27}  \, n_f^{2}l_{q m}
-\frac{3251}{1152}  \,\zeta_{3}l_{q m}
  \right. \nonumber \\ &{}& \left.  
\phantom{+z^{2}}
-\frac{1303}{24} l_{q \mu}l_{q m}
+\frac{887}{432}  \,n_f l_{q \mu}l_{q m}
+\frac{1}{36}  \, n_f^{2}l_{q \mu}l_{q m}
-\frac{773}{288} l_{q \mu}^{2}l_{q m}
  \right. \nonumber \\ &{}& \left.  
\phantom{+z^{2}}
-\frac{49}{72}  \,n_f l_{q \mu}^{2}l_{q m}
-\frac{2431}{192} l_{q m}^{2}
+\frac{1271}{1728}  \,n_f l_{q m}^{2}
-\frac{6947}{576} l_{q \mu}l_{q m}^{2}
  \right. \nonumber \\ &{}& \left.  
\phantom{+z^{2}}
-\frac{13121}{3456} l_{q m}^{3}
+\frac{49}{864}  \,n_f l_{q m}^{3}
\right]
{}.
\label{SigmaS3M}
\end{eqnarray}

\section{$\mbox{RI}$ scheme versus $\overline{\mbox{MS}}$ and
$\mbox{MOM}$ schemes}
\label{sec:RIvsMSandMOM}

\subsection{Three Loop Conversion functions}
\label{sub:conversion3l}

A direct use of
Eqs.~(\ref{form:C2RI},\ref{form:CmRI},\ref{form:C2RIp}) and
(\ref{form:CmRIp}) leads to the following analytical expressions for
the conversion factors between the $ \overline{\mathrm{MS}} $ and $
\mathrm{RI} $ schemes. The results are shown for QCD ($ \mathrm{SU}
(3)$) and Landau gauge as functions of $n_f$ ( the $\zeta_i$ are the
values $\zeta(i)$ of Riemann's Zeta function ):

\begin{eqnarray}C_2^{ \mathrm{RI} } = & & 
1 \quad + \quad \left(\frac{\alpha_s}{4\pi}\right)^{2}
\left[
-\frac{517}{18} 
+12  \,\zeta_{3}
+\frac{5}{3}  \,n_f 
\right]
\nonumber\\
 &{+}& \left(\frac{\alpha_s}{4\pi}\right)^{3}
\left[
-\frac{1287283}{648} 
+\frac{14197}{12}  \,\zeta_{3}
+\frac{79}{4}  \,\zeta_{4}
-\frac{1165}{3}  \,\zeta_{5}
  \right. \nonumber \\ &{}& \left.  
\phantom{+\left(\frac{\alpha_s}{4\pi}\right)^{3}}
+\frac{18014}{81}  \,n_f 
-\frac{368}{9}  \,\zeta_{3} \,n_f 
-\frac{1102}{243}  \, n_f^{2}
\right]
{},
\label{C2RIqcd}
\end{eqnarray}

\begin{eqnarray}C_m^{ \mathrm{RI} } = & & 
1 \quad + \quad  \frac{\alpha_s}{4\pi}
\left[ 
-\frac{16}{3}\right] 
 \quad + \quad \left(\frac{\alpha_s}{4\pi}\right)^{2}
\left[
-\frac{1990}{9} 
+\frac{152}{3}  \,\zeta_{3}
+\frac{89}{9}  \,n_f 
\right]
\nonumber\\
 &{+}& \left(\frac{\alpha_s}{4\pi}\right)^{3}
\left[
-\frac{6663911}{648} 
+\frac{408007}{108}  \,\zeta_{3}
-\frac{2960}{9}  \,\zeta_{5}
+\frac{236650}{243}  \,n_f 
  \right. \nonumber \\ &{}& \left.  
\phantom{+\left(\frac{\alpha_s}{4\pi}\right)^{3}}
-\frac{4936}{27}  \,\zeta_{3} \,n_f 
+\frac{80}{3}  \,\zeta_{4} \,n_f 
-\frac{8918}{729}  \, n_f^{2}
-\frac{32}{27}  \,\zeta_{3} \, n_f^{2}
\right]
{},
\label{CmRIqcd}
\end{eqnarray}

\begin{eqnarray}C_2^{ \mathrm{RI'} } = & & 
1 \quad + \quad \left(\frac{\alpha_s}{4\pi}\right)^{2}
\left[
-\frac{359}{9} 
+12  \,\zeta_{3}
+\frac{7}{3}  \,n_f 
\right]
\nonumber\\
 &{+}& \left(\frac{\alpha_s}{4\pi}\right)^{3}
\left[
-\frac{439543}{162} 
+\frac{8009}{6}  \,\zeta_{3}
+\frac{79}{4}  \,\zeta_{4}
-\frac{1165}{3}  \,\zeta_{5}
  \right. \nonumber \\ &{}& \left.  
\phantom{+\left(\frac{\alpha_s}{4\pi}\right)^{3}}
+\frac{24722}{81}  \,n_f 
-\frac{440}{9}  \,\zeta_{3} \,n_f 
-\frac{1570}{243}  \, n_f^{2}
\right]
{},
\label{C2RIpqcd}
\end{eqnarray}

\begin{eqnarray}C_m^{ \mathrm{RI'} } = & & 
1 \quad + \quad  \frac{\alpha_s}{4\pi}
\left[ 
-\frac{16}{3}\right] 
 \quad + \quad \left(\frac{\alpha_s}{4\pi}\right)^{2}
\left[
-\frac{3779}{18} 
+\frac{152}{3}  \,\zeta_{3}
+\frac{83}{9}  \,n_f 
\right]
\nonumber\\
 &{+}& \left(\frac{\alpha_s}{4\pi}\right)^{3}
\left[
-\frac{3115807}{324} 
+\frac{195809}{54}  \,\zeta_{3}
-\frac{2960}{9}  \,\zeta_{5}
+\frac{217390}{243}  \,n_f 
  \right. \nonumber \\ &{}& \left.  
\phantom{+\left(\frac{\alpha_s}{4\pi}\right)^{3}}
-\frac{4720}{27}  \,\zeta_{3} \,n_f 
+\frac{80}{3}  \,\zeta_{4} \,n_f 
-\frac{7514}{729}  \, n_f^{2}
-\frac{32}{27}  \,\zeta_{3} \, n_f^{2}
\right]
{}.
\label{CmRIpqcd}
\end{eqnarray}

At a scale of $\mu = 2$ GeV and $n_f=4$, the numerical contributions
of the leading order to NNNLO terms are as follows (for simplicity we
inserted $\alpha_s/\pi = 0.1$):
\begin{equation}
C_2^{ \mathrm{RI} } = 1.0 + 0.0 - 0.00476 - 0.00508 \, {},
\label{C2RInum}
\end{equation}
\begin{equation}
C_m^{ \mathrm{RI} } = 1.0 - 0.1333 - 0.0754 - 0.0495  
\label{CmRInum}
\end{equation}
and
\begin{equation}
C_2^{ \mathrm{RI'} } = 1.0 + 0.0 - 0.0101 - 0.0095 \, {},
\label{C2RIpnum}
\end{equation}
\begin{equation}
C_m^{ \mathrm{RI'} } = 1.0 - 0.1333 - 0.0701 - 0.0458 \, {}.
\label{CmRIpnum}
\end{equation}
One observes that the sizes of the NNLO and NNNLO contributions to
$C_m^{ \mathrm{RI} }$ at this scale amount to about 7.5\% and 5\%
respectively. This shows that perturbation theory can not be used for
a precise conversion of the $ \mathrm{RI} $ quark masses to the $
\overline{\mathrm{MS}} $ ones at the renormalization scale $\mu = $ 2
GeV. The convergence can be improved if one increases $\mu$ to, say, 3
GeV.  Indeed, with this choice of $\mu$ the standard three-loop
evolution gives $\alpha_s(3 \mbox{ GeV}) = 0.262$ and
Eqs.~(\ref{C2RInum},\ref{CmRInum}) transform to
\begin{equation}
C_2^{ \mathrm{RI} } = 1.0 + 0.0 - 0.00333 - 0.00296 
\label{C2_3}
{}
\end{equation}
and
\begin{equation}
C_m^{ \mathrm{RI} } =  1.0 - 0.111 - 0.0526 - 0.0289 
{}.
\label{Cm_3}
\end{equation}

The accuracy of the massless approximation can be tested by computing
the ratio $C_?^{ \mathrm{RI} }/C_?^{ \mathrm{MOM} }\quad (? = m,2)$ as
a series in $z = m_q^2/(-\mu^2)$.  Using the results of the previous
section one obtains

\begin{eqnarray}
\frac{C_2^{ \mathrm{RI} }}{C_2^{ \mathrm{MOM} }} = & & 
1\nonumber\\
 &{+}&  a_s^{2} 
\left[
0.28981\, z
-0.89236\, z^{2}
+1.5284\, z^{3}
-3.3649\, z^{4}
+7.6945\, z^{5}
  \right. \nonumber \\ &{}& \left.  
\phantom{+ a_s^{2} }
+0.25\, zl_z
-0.39583\, z^{2}l_z
+0.45602\, z^{3}l_z
-2.2796\, z^{4}l_z
  \right. \nonumber \\ &{}& \left.  
\phantom{+ a_s^{2} }
+23.231\, z^{5}l_z
+0.024306\, z^{3}l_z^{2}
+1.1007\, z^{4}l_z^{2}
-8.8726\, z^{5}l_z^{2}
\right]
\nonumber\\
 &{+}&  a_s^{3} 
\left[
4.4496\, z
-20.979\, z^{2}
+2.5931\, zl_z
-3.0558\, z^{2}l_z
  \right. \nonumber \\ &{}& \left.  
\phantom{+ a_s^{3} }
+0.70052\, zl_z^{2}
-0.49414\, z^{2}l_z^{2}
\right]
{},
\label{C2RIbyCMOM}
\end{eqnarray}

\begin{eqnarray}
\frac{C_m^{ \mathrm{RI} }}{C_m^{ \mathrm{MOM} }} = & & 
1\nonumber\\
 &{+}&  a_s 
\left[
-1.0\, z
-0.5\, z^{2}
+0.16667\, z^{3}
-0.083333\, z^{4}
+0.05\, z^{5}
 \right. \nonumber \\ &{}& \left.  
\phantom{+ a_s^{2} }
-1.0\, zl_z
\right]
\nonumber\\
 &{+}&  a_s^{2} 
\left[
-7.6458\, z
+0.86458\, z^{2}
-3.5129\, z^{3}
+13.913\, z^{4}
-36.828\, z^{5}
  \right. \nonumber \\ &{}& \left.  
\phantom{+ a_s^{2} }
-6.3264\, zl_z
+0.10417\, z^{2}l_z
+2.3866\, z^{3}l_z
+7.3259\, z^{4}l_z
  \right. \nonumber \\ &{}& \left.  
\phantom{+ a_s^{2} }
-71.347\, z^{5}l_z
-2.0\, zl_z^{2}
+1.0208\, z^{2}l_z^{2}
+1.1956\, z^{3}l_z^{2}
  \right. \nonumber \\ &{}& \left.  
\phantom{+ a_s^{2} }
-6.706\, z^{4}l_z^{2}
+30.991\, z^{5}l_z^{2}
\right]
\nonumber\\
 &{+}&  a_s^{3} 
\left[
-59.008\, z
+48.743\, z^{2}
-41.464\, zl_z
-5.4235\, z^{2}l_z
  \right. \nonumber \\ &{}& \left.  
\phantom{+ a_s^{3} }
-18.829\, zl_z^{2}
+9.1024\, z^{2}l_z^{2}
-4.0556\, zl_z^{3}
+3.5697\, z^{2}l_z^{3}
\right]
{},
\label{CmRIbyCMOM}
\end{eqnarray}
where $l_z = log(-m^2/\mu^2)$ and we have evaluated the coefficients
in the series in $z$ with $n_f=4$.

\begin{figure}
\begin{center}
\epsfig{file=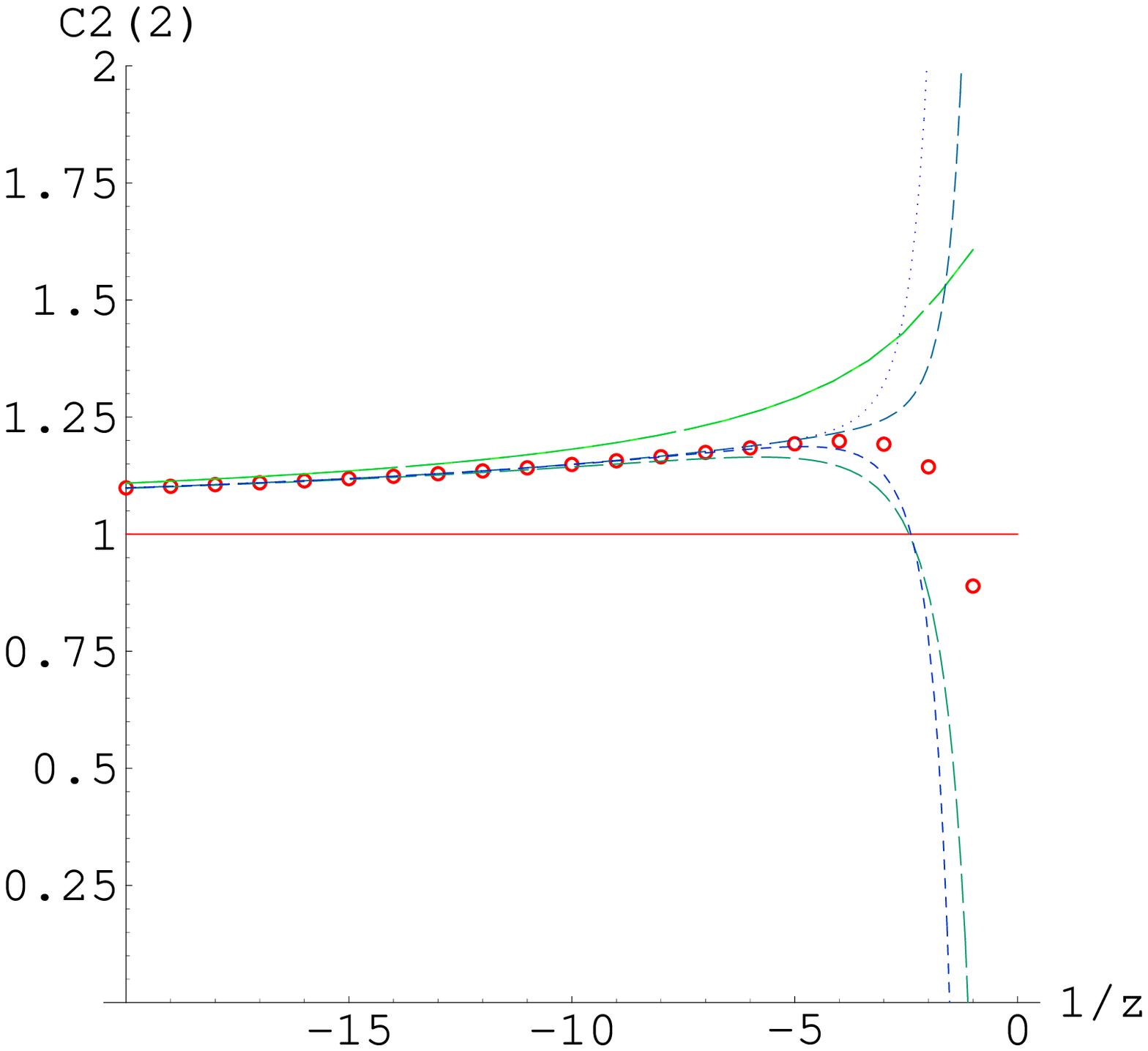,height=10.0cm,width=6.0cm}
\epsfig{file=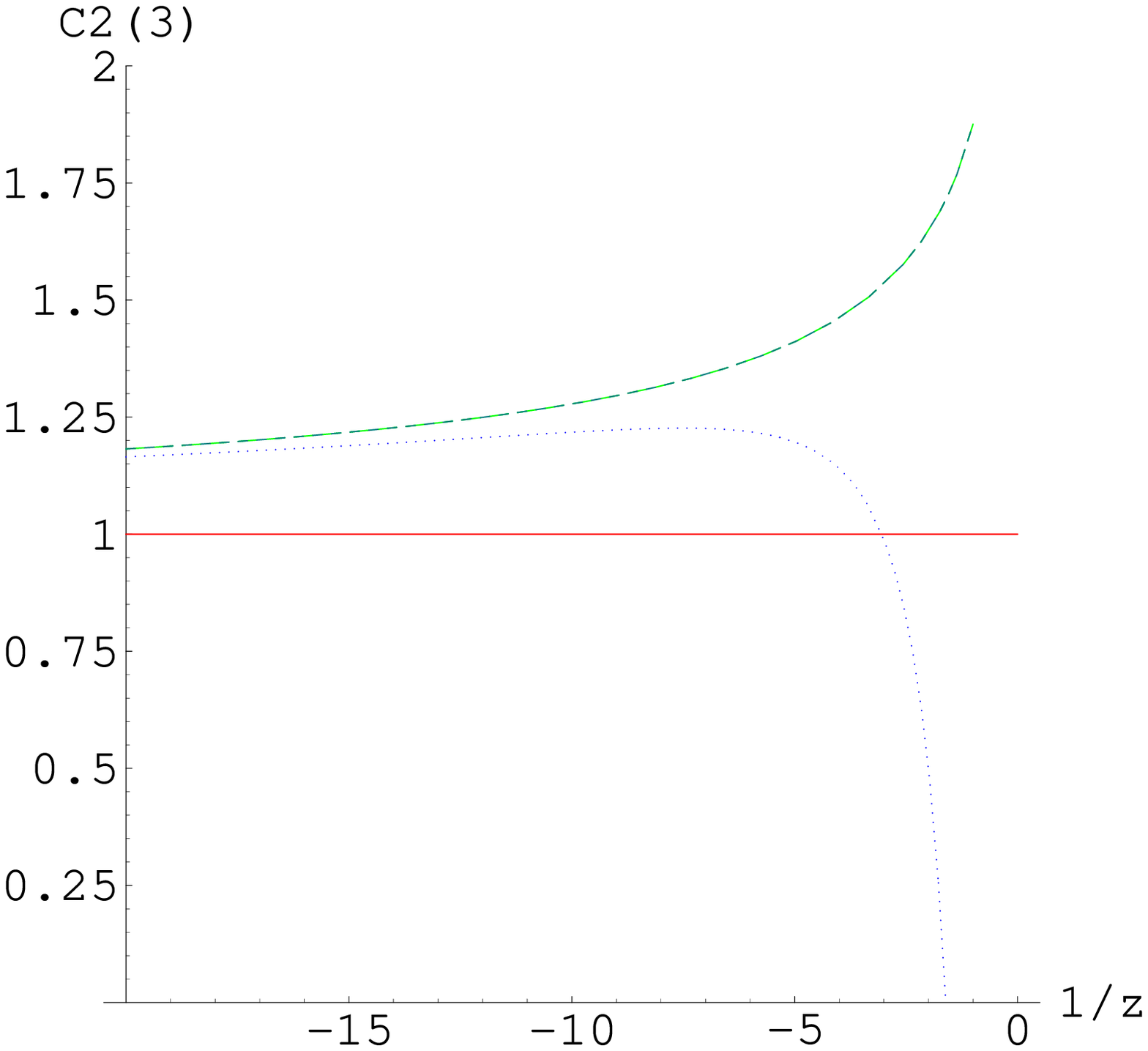,height=10.0cm,width=6.0cm}
\end{center}
\caption{The ratio of the $ \mathrm{RI} $ and $ \mathrm{MOM} $ scheme
conversion functions $C_{2,l}^{ \mathrm{MOM} }/C_{2,l}^{ \mathrm{RI}
}$ as functions of $-\mu^2/m^2$. Shown are the coefficients in the
expansion in $a_s/\pi$ as expansions to order $z$ to $z^5$ ($z^2$ for
3 loops). Note that in Landau gauge $C_2^? = 0$ , for 2 loops some
numeric values for the exact mass dependence are shown as well.}
\label{plot:C2}
\end{figure}

\begin{figure}
\begin{center}
\epsfig{file=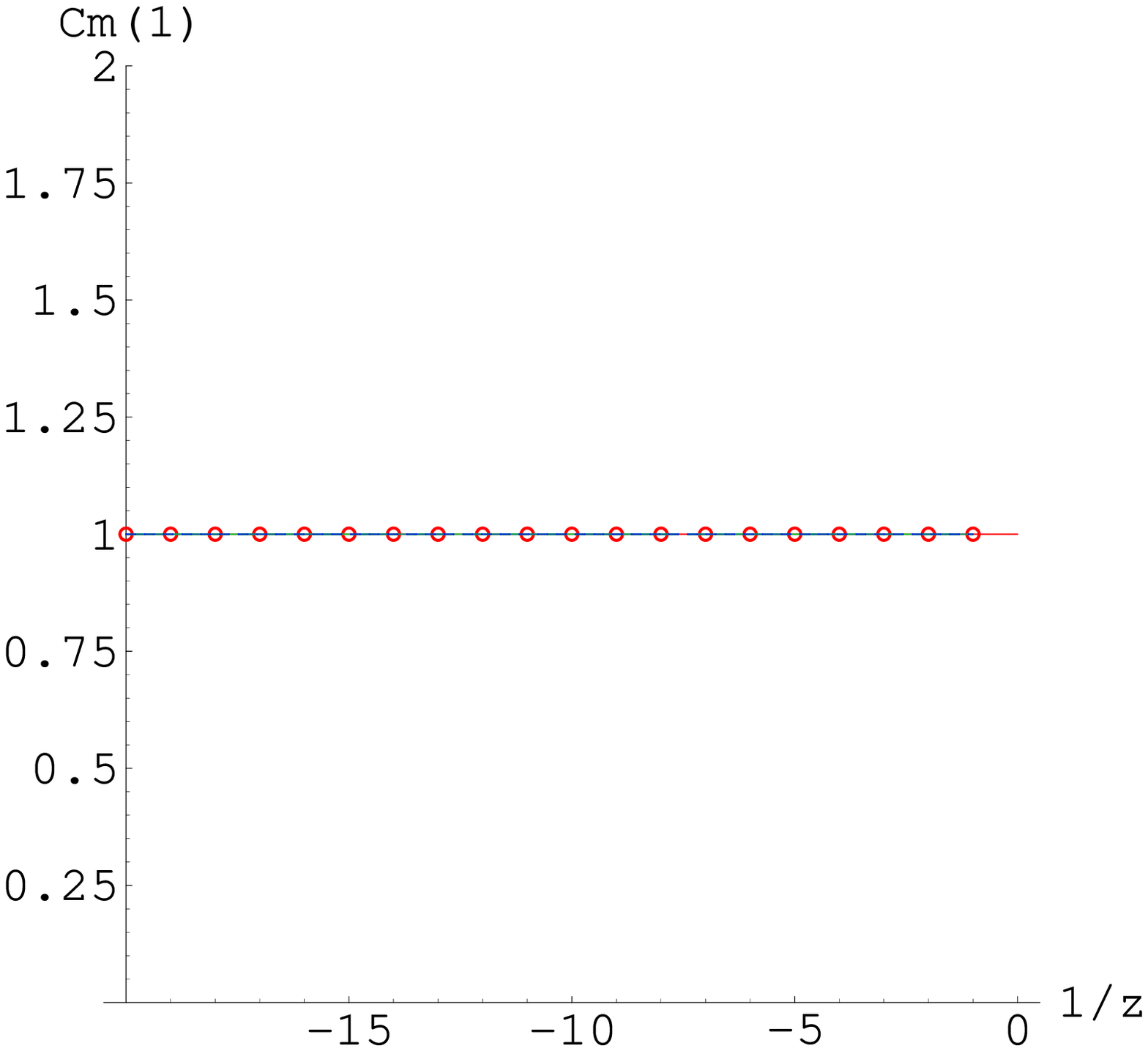,height=10.0cm,width=6.0cm}
\epsfig{file=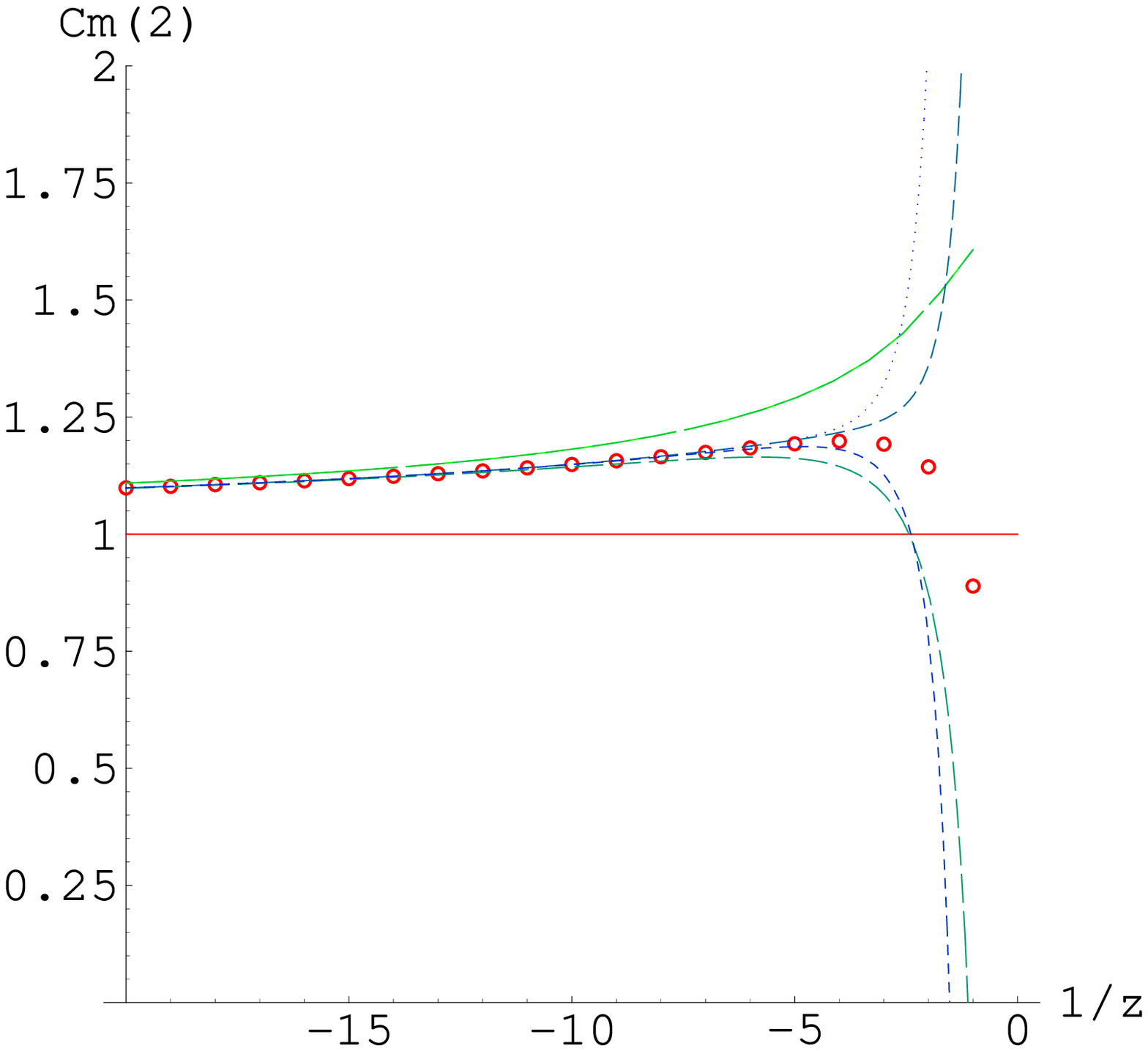,height=10.0cm,width=6.0cm}

\vspace{-1.8cm}

\epsfig{file=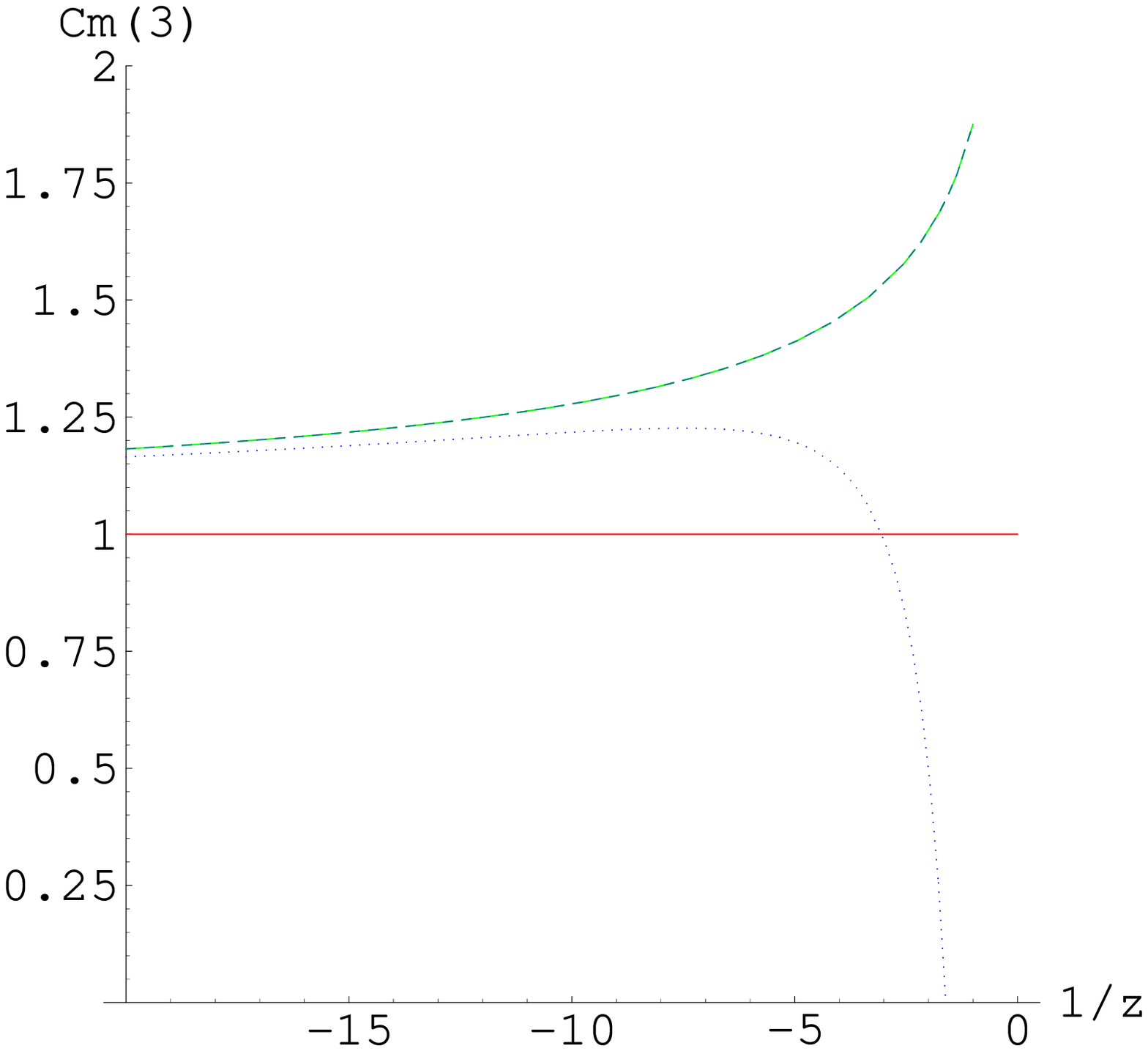,height=10.0cm,width=6.0cm}
\end{center}
\caption{The ratio of the $ \mathrm{RI} $ and $ \mathrm{MOM} $ scheme
conversion functions $C_{m,l}^{ \mathrm{MOM} }/C_{m,l}^{ \mathrm{RI}
}$ as functions of $-\mu^2/m^2$. Shown are the coefficients in the
expansion in $a_s/\pi$ as expansions to order $z$ to $z^5$ ($z^2$ for
3 loops). For 1 and 2 loops some numeric values for the exact mass
dependence are shown as well.}
\label{plot:Cm}
\end{figure}

To illustrate the quality of these expansions, we have plotted (see
figures \ref{plot:C2} and \ref{plot:Cm}) the ratio of the 1, 2 and 3
loop coefficients of $C_m^{ \mathrm{MOM} }$ and $C_m^{ \mathrm{RI} }$
as functions of $1/z=-\mu^2/m^2$ in the Landau gauge and for
simplicity with $n_f=4$ for all values of $z$. The circles in the
plots for 1 and 2 loops correspond to the exact results from
\cite{Fleischer:1998dw}. The convergence of the small mass
(corresponding to large negative values of $1/z$) expansions is good
for $1/z < -4$, where the expansions for higher orders of $z$ are
almost indistinguishable as well among each other as well as from the
numbers received from the exact 2 loop propagator. On the other hand,
due to the $z\,\ln(z)^i , i=1,\dots,l$ ( where $l$ is the number of
loops) terms, the $ \mathrm{MOM} $ coefficients are approaching the
corresponding values in the $ \mathrm{RI} $-scheme for increasing
$1/|z|$ only very slowly. This makes the $ \mathrm{RI} $-scheme as an
approximation to the $ \mathrm{MOM} $-scheme for the c quark useless.

\subsection{Four Loop Quark Anomalous Dimensions}
\label{sub:anomaldims4l}

We start from the $ \overline{\mathrm{MS}} $ scheme. The quark mass
anomalous dimension was computed at four loops quite recently in
~\cite{Chetyrkin:1997dh,Vermaseren:1997fq} and reads
\[
\gamma_m(a_s) \equiv                                           
-\sum_{i\geq0}\gamma_m^{(i)}                                         
a_s^{i+1}                                                             
{},        
\]

\begin{eqnarray}\gamma_{m}^{(0)} = & & 
1{}
\label{MassAnomalDimMS0qcd}
\end{eqnarray}

\begin{eqnarray}\gamma_{m}^{(1)} = \frac{1}{16}\Bigg\{ & & 
 \frac{202}{3} \, + \,  \,n_f 
\left[ 
-\frac{20}{9}\right] 
\,\Bigg\}{}
\label{MassAnomalDimMS1qcd}
\end{eqnarray}

\begin{eqnarray}\gamma_{m}^{(2)} = \frac{1}{64}\Bigg\{ & & 
 1249 +  \,n_f 
\left[
-\frac{2216}{27} 
-\frac{160}{3}  \,\zeta_{3}
\right]
 +  \, n_f^{2}
\left[ 
-\frac{140}{81}\right] 
\,\Bigg\}{}
\label{MassAnomalDimMS2qcd}
\end{eqnarray}

\begin{eqnarray}\gamma_{m}^{(3)} = \frac{1}{256}\Bigg\{ & & 
\left[
\frac{4603055}{162} 
+\frac{135680}{27}  \,\zeta_{3}
-8800  \,\zeta_{5}
\right]
\nonumber\\
 &{+}&  \,n_f 
\left[
-\frac{91723}{27} 
-\frac{34192}{9}  \,\zeta_{3}
+880  \,\zeta_{4}
+\frac{18400}{9}  \,\zeta_{5}
\right]
\nonumber\\
 &{+}&  \, n_f^{2}
\left[
\frac{5242}{243} 
+\frac{800}{9}  \,\zeta_{3}
-\frac{160}{3}  \,\zeta_{4}
\right]
 \, + \,  \, n_f^{3}
\left[
-\frac{332}{243} 
+\frac{64}{27}  \,\zeta_{3}
\right]
\,\Bigg\}{}.
\label{MassAnomalDimMS3qcd}
\end{eqnarray}

The result for the quark field anomalous dimension was found by one of
the authors in the course of computing $\gamma_m$
\cite{Chetyrkin:1997dh} and read for the QCD case in Landau
gauge\footnote{$\gamma_{2}$ is gauge dependent and in the Landau gauge
$\gamma_{2}^{(0)} = 0$; the results for $ \mathrm{SU} (N)$ group in
general covariant gauge are given in Appendix C.}
\[
\gamma_2(a_s) \equiv                                           
-\sum_{i\geq0}\gamma_2^{(i)}                                         
a_s^{i+1}                                                             
{},        
\]

\begin{eqnarray}\gamma_{2}^{(1)} = \frac{1}{16}\Bigg\{ & & 
 \frac{67}{3} \, + \,  \,n_f 
\left[ 
-\frac{4}{3}\right] 
\,\Bigg\}{}
\label{FieldAnomalDimMS1qcd}
\end{eqnarray}

\begin{eqnarray}\gamma_{2}^{(2)} = \frac{1}{64}\Bigg\{ & & 
\left[
\frac{20729}{36} 
-\frac{79}{2}  \,\zeta_{3}
\right]
 +  \,n_f 
\left[ 
-\frac{550}{9}\right] 
 +  \, n_f^{2}
\left[ 
 \frac{20}{27}\right] 
\,\Bigg\}{}
\label{FieldAnomalDimMS2qcd}
\end{eqnarray}

\begin{eqnarray}\gamma_{2}^{(3)} = \frac{1}{256}\Bigg\{ & & 
\left[
\frac{2109389}{162} 
-\frac{565939}{324}  \,\zeta_{3}
+\frac{2607}{4}  \,\zeta_{4}
-\frac{761525}{1296}  \,\zeta_{5}
\right]
\nonumber\\
 &{+}&  \,n_f 
\left[
-\frac{162103}{81} 
-\frac{2291}{27}  \,\zeta_{3}
-\frac{79}{2}  \,\zeta_{4}
-\frac{160}{3}  \,\zeta_{5}
\right]
\nonumber\\
 &{+}&  \, n_f^{2}
\left[
\frac{3853}{81} 
+\frac{160}{9}  \,\zeta_{3}
\right]
 \, + \,  \, n_f^{3}
\left[ 
 \frac{140}{243}\right] 
\,\Bigg\}{}.
\label{FieldAnomalDimMS3qcd}
\end{eqnarray}

For completeness we also give the field anomalous dimension for the
case of QED with $n_f$ different fermion species\footnote{For the QED
case, all gauge dependence is in $\gamma_{2}^{(0)}$, where $\xi_L$ is
defined in Appendix \ref{app:quarkprop3l} ($\xi_L=0$ for Landau
gauge)}:

\begin{eqnarray}\gamma_{2}^{QED\,(0)} = & & \frac{1}{4}
\left[
  \,\xi_{L} 
\right]
{}
\label{FieldAnomalDimMS0qed}
\end{eqnarray}

\begin{eqnarray}\gamma_{2}^{QED\,(1)} = \frac{1}{16}\Bigg\{ & & 
-\frac{3}{2} \, + \,  \,n_f 
\left[ 
-2\right] 
\,\Bigg\}
\label{FieldAnomalDimMS1qed}
\end{eqnarray}

\begin{eqnarray}\gamma_{2}^{QED\,(2)} = \frac{1}{64}\Bigg\{ & & 
 \frac{3}{2} +  \,n_f 
\left[ 
 3\right] 
 +  \, n_f^{2}
\left[ 
 \frac{20}{9}\right] 
\,\Bigg\}
\label{FieldAnomalDimMS2qed}
\end{eqnarray}

\begin{eqnarray}\gamma_{2}^{QED\,(3)} = \frac{1}{256}\Bigg\{ & & 
\left[
-\frac{1027}{8} 
-400  \,\zeta_{3}
+640  \,\zeta_{5}
\right]
 \, + \,  \,n_f 
\left[
\frac{460}{3} 
-64  \,\zeta_{3}
\right]
\nonumber\\
 &{+}&  \, n_f^{2}
\left[
\frac{304}{9} 
-32  \,\zeta_{3}
\right]
 \, + \,  \, n_f^{3}
\left[ 
 \frac{280}{81}\right] 
\,\Bigg\}{}.
\label{FieldAnomalDimMS3qed}
\end{eqnarray}

In order to compute the corresponding anomalous dimensions for the $
\mathrm{RI} $ and $ \mathrm{RI'} $ schemes, one just needs to make use
of Eqs.~(\ref{MassAnomalDimMS0qcd}) to (\ref{MassAnomalDimMS3qcd}) in
combination with the three loop conversion functions
(\ref{C2RIqcd},\ref{CmRIpqcd}).  As a result for the QCD case we get
for the $ \mathrm{RI} $ scheme:

\begin{eqnarray}\gamma_{m}^{ \mathrm{RI}  (0)} = & & 
1 
\label{MassAnomalDimRI0qcd}
\end{eqnarray}

\begin{eqnarray}\gamma_{m}^{ \mathrm{RI}  (1)} = \frac{1}{16}\Bigg\{ & & 
 126 +  \,n_f 
\left[ 
-\frac{52}{9}\right] 
\,\Bigg\}
\label{MassAnomalDimRI1qcd}
\end{eqnarray}

\begin{eqnarray}\gamma_{m}^{ \mathrm{RI}  (2)} = \frac{1}{64}\Bigg\{ & & 
\left[
\frac{20911}{3} 
-\frac{3344}{3}  \,\zeta_{3}
\right]
 +  \,n_f 
\left[
-\frac{18386}{27} 
+\frac{128}{9}  \,\zeta_{3}
\right]
\nonumber\\
 &{+}&  \, n_f^{2}
\left[ 
 \frac{928}{81}\right] 
\,\Bigg\}\nonumber \\
\label{MassAnomalDimRI2qcd}
\end{eqnarray}

\begin{eqnarray}\gamma_{m}^{ \mathrm{RI}  (3)} = \frac{1}{256}\Bigg\{ & & 
\left[
\frac{300665987}{648} 
-\frac{15000871}{108}  \,\zeta_{3}
+\frac{6160}{3}  \,\zeta_{5}
\right]
\nonumber\\
 &{+}&  \,n_f 
\left[
-\frac{7535473}{108} 
+\frac{627127}{54}  \,\zeta_{3}
+\frac{4160}{3}  \,\zeta_{5}
\right]
\nonumber\\
 &{+}&  \, n_f^{2}
\left[
\frac{670948}{243} 
-\frac{6416}{27}  \,\zeta_{3}
\right]
 \, + \,  \, n_f^{3}
\left[ 
-\frac{18832}{729}\right] 
\,\Bigg\}{}.
\label{MassAnomalDimRI3qcd}
\end{eqnarray}

The corresponding equations for the $ \mathrm{RI'} $ scheme are:

\begin{eqnarray}\gamma_{m}^{ \mathrm{RI'}  (0)} = & & 
1
\label{MassAnomalDimRIp0qcd}
\end{eqnarray}

\begin{eqnarray}\gamma_{m}^{ \mathrm{RI'}  (1)} = \frac{1}{16}\Bigg\{ & & 
 126 +  \,n_f 
\left[ 
-\frac{52}{9}\right] 
\,\Bigg\}
\label{MassAnomalDimRIp1qcd}
\end{eqnarray}

\begin{eqnarray}\gamma_{m}^{ \mathrm{RI'}  (2)} = \frac{1}{64}\Bigg\{ & & 
\left[
\frac{20174}{3} 
-\frac{3344}{3}  \,\zeta_{3}
\right]
 +  \,n_f 
\left[
-\frac{17588}{27} 
+\frac{128}{9}  \,\zeta_{3}
\right]
\nonumber\\
 &{+}&  \, n_f^{2}
\left[ 
 \frac{856}{81}\right] 
\,\Bigg\}\nonumber \\
\label{MassAnomalDimRIp2qcd}
\end{eqnarray}

\begin{eqnarray}\gamma_{m}^{ \mathrm{RI'}  (3)} = \frac{1}{256}\Bigg\{ & & 
\left[
\frac{141825253}{324} 
-\frac{7230017}{54}  \,\zeta_{3}
+\frac{6160}{3}  \,\zeta_{5}
\right]
\nonumber\\
 &{+}&  \,n_f 
\left[
-\frac{3519059}{54} 
+\frac{298241}{27}  \,\zeta_{3}
+\frac{4160}{3}  \,\zeta_{5}
\right]
\nonumber\\
 &{+}&  \, n_f^{2}
\left[
\frac{611152}{243} 
-\frac{5984}{27}  \,\zeta_{3}
\right]
 \, + \,  \, n_f^{3}
\left[ 
-\frac{16024}{729}\right] 
\,\Bigg\}{}.
\label{MassAnomalDimRIp3qcd}
\end{eqnarray}

The quark field anomalous dimensions for the $ \mathrm{RI} $ and $
\mathrm{RI'} $ schemes are given in Appendix C.

Common to all our results for the quark mass and field anomalous
dimensions in the RI and RI' schemes (see
Eqs.~(\ref{MassAnomalDimRIp3qcd}) and Appendix C) is the absence of
$\zeta_4$ even at the four-loop level.  This is in striking contrast
to the behavior of their $\overline{\mathrm{MS}}$ counterparts and
calls for an explanation. In fact, a similar absence of the constant
$\zeta_4$ at four loops in the gauge beta functions recently has been
neatly explained in \cite{Broadhurst:1999xk}.

Unfortunately, we have not been able to extend the argument of
\cite{Broadhurst:1999xk} to the present case. Nevertheless, we tend to
agree with David Broadhurst and the referee in suggesting that the
appearance of $\zeta_4$ in
Eqs.~(\ref{MassAnomalDimMS3qcd},\ref{FieldAnomalDimMS3qcd}) is
apparently an artifact of the $\overline{\mathrm{MS}}$-scheme.

To support this idea we demonstrate below that, in a sense, the RI/RI'
schemes are "more physical" than the $\overline{\mathrm{MS}}$ one. To
illustrate this statement, let us define the $q$ dependent
generalization --- $\hat{C}_{?}^{RI}(\alpha_s,\mu^2/-q^2)$, with $? =
2$ or $m$, --- of the conversion functions $\hat{C}_{?}^{RI}$ by the
same Eqs.~(\ref{form:C2RI}-\ref{form:CmRI}) but {\em without } the
condition $q^2=-\mu^2$.  Let us in addition define
\begin{eqnarray}
\Gamma_?^{RI}(\alpha_s,\mu^2/-q^2) &=&  \mu^2\frac{\partial }{\partial \mu^2}
\log \left[
\hat{C}_{?}^{RI}(\alpha_s,\mu^2/-q^2)
\right]
\label{Gamma_mu}
\\
{} 
&=&  -q^2\frac{\partial }{\partial q^2}
\log \left[
\hat{C}_{?}^{RI}(\alpha_s,\mu^2/-q^2)
\right]
\label{Gamma_q}
{}. 
\end{eqnarray}

Using these definitions, the following relations hold:
\begin{itemize}
\item Boundary condition: 
\[
\hat{C}_{?}^{RI}(\alpha_s,1) = {C}_{?}^{RI}(\alpha_s)
{}. 
\]
\item 
Evolution equations (they follow directly from the evolution
equation of the fermion propagator):
\begin{equation}
\mu^2\frac{\mathrm{d}}{\mathrm{d}\mu^2} {\hat C}_{?}^{RI} = 
\gamma_? {\hat C}_{?}^{RI}
, \qquad  \mu^2\frac{\mathrm{d}}{\mathrm{d}\mu^2} 
\Gamma_?^{RI} =0
\label{RG:Gamma:1}
\end{equation}
or, equivalently, 
\begin{equation}
(\mu^2 \frac{\partial  }{\partial \mu^2}
+
\beta\frac{\partial}{\partial a_s }){\hat C}_{?}^{RI} = 
\gamma_? {\hat C}_{?}^{RI}
{},\qquad
(\mu^2 \frac{\partial  }{\partial \mu^2}
+
\beta\frac{\partial}{\partial a_s }) \Gamma_?^{RI} =0
\label{RG:Gamma:2}
{}.
\end{equation}
\item 
Relations between the $\Gamma_?^{RI}$ and the $\gamma_?^{RI}$ (can be
received by combining Eqs.~(\ref{RG:Gamma:1},\ref{RG:Gamma:2}) and
(\ref{form:gamma-pm},\ref{form:gamma-p2}):
\begin{equation}
\gamma_?^{RI} (\alpha_s) = \Gamma_?^{RI}(\alpha_s,1)
\label{gamma=Gamma}
{}.
\end{equation}
\end{itemize}

In fact, the functions $\Gamma_?^{RI}$ happen to be both scale {\em
and} scheme independent. In Eqs.~(\ref{form:C2RI}-\ref{form:CmRI}) it
is understood that the fermion propagator is defined in the
$\overline{\mathrm{MS}}$-scheme.  However, one can easily check that
even if the fermion propagator would be defined in any other
(mass-independent) scheme, then the resulting change of the
$\hat{C}_{?}^{RI}$ would amount to a rescaling by a $q$-independent
factor which can not, obviously, change the very functions
$\Gamma_?^{RI}$.

We conclude that the $\Gamma_?^{RI}$ are physical scheme invariant
quantities, at least within the framework of perturbation
theory. Thus, due to the relation (\ref{gamma=Gamma}), the absence of
$\zeta_4$ in the quark mass and field anomalous dimensions in the $RI$
scheme should be considered as a phenomenon which is no more puzzling
than the similar absence\footnote{We do not count the well-understood
$\pi^2$-term arising due to the analytical continuation to the
physical region of energies.} of $\zeta_4$ in the four-loop total
cross-section of $e^+e^-$ annihilation into hadrons calculated in
massless QCD \cite{GorKatLar91SurSam91,gvvq}.

The same reasoning is fully applicable to the case of the $RI'$
scheme.

\subsection{NNNLO relation for the $ \mathrm{RI} $ quark mass and the 
$\mathrm{RGI}$ mass $\hat{m}_q$}\label{sub:RIandRGmasses}

It is customary to solve the RG equation (\ref{def:anomal-mass}) for
the quark running mass $m(\mu)$ as follows
\begin{equation} 
\frac{m(\mu)}{m(\mu_0)} = \frac{c(a_s(\mu))}{c(a_s(\mu_0))}
\label{def:c-functions}
{},
\end{equation}
where ($x$ stands for either $\alpha_s(\mu)/\pi$ or $\alpha_s(\mu_0)/
\pi$)
\begin{eqnarray} 
c(x) &=& 
\exp \,\left\{
\int^{x}
     {\rm d} x' \frac{\gamma_m(x')}{\beta(x')}\right\}
\nonumber \\ 
 &=&
(x)^{\bar{\gamma_0}}  \left\{ 1 + (\bar{\gamma_1} - \bar{\beta_1}\bar{\gamma_0})x
\right.
\nonumber \\ 
&+&
\frac{1}{2}
\left[
(\bar{\gamma_1} - \bar{\beta_1}\bar{\gamma_0})^2 
+
\bar{\gamma_2} + \bar{\beta_1}^2\bar{\gamma_0}
- \bar{\beta_1}\bar{\gamma_1} -\bar{\beta_2}\bar{\gamma_0}
\right] x^2
\nonumber \\ 
&+&
\left[ 
\frac{1}{6}(\bar{\gamma_1} - \bar{\beta_1}\bar{\gamma_0})^3
+
\frac{1}{2}(\bar{\gamma_1} - \bar{\beta_1}\bar{\gamma_0})
(
\bar{\gamma_2} + \bar{\beta_1}^2\bar{\gamma_0}
- \bar{\beta_1}\bar{\gamma_1} -\bar{\beta_2}\bar{\gamma_0}
)
\right.
\nonumber \\ 
&&
+\frac{1}{3}\left(
\left.\left.
\bar{\gamma_3}
-\bar{\beta_1}^3\bar{\gamma_0} + 2\bar{\beta_1} \bar{\beta_2}\bar{\gamma_0}
-\bar{\beta_3}\bar{\gamma_0} + \bar{\beta_1}^2\bar{\gamma_1}
- \bar{\beta_2}\bar{\gamma_1} - \bar{\beta_1}\bar{\gamma_2}
\right)
\right] x^3 \right. 
\nonumber \\
& & + \left. {
O}(x^4)
\right\}
{}. \label{c(x)}
\end{eqnarray}

Here $\bar{\gamma_i} = \gamma_m^{(i)}/\beta_0$, $\bar{\beta_i} =
\beta_i/\beta_0$, ($i$=1,2,3) and $\beta_i$ are the coefficients of
the QCD beta-function as defined in Eq.~(\ref{def:beta}).  The
four-loop beta-function recently has been computed in
\cite{vanRitbergen:1997va} with the result
\begin{equation}\label{form:beta}
\begin{array}{ll}\displaystyle
\beta_0 = &\displaystyle\frac{1}{4}\left(11-\frac{2}{3}n_f\right),
  \;\;\;\;  
\beta_1=\frac{1}{16}\left(102-\frac{38}{3}n_f\right), 
\\[3ex] \displaystyle
\beta_2  = & \displaystyle \frac{1}{64}\left(\frac{2857}{2}
-\frac{5033}{18}n_f+ 
\frac{325}{54}n_f^2\right),
\\[3ex]
\beta_3  = &  \displaystyle \frac{1}{256}\left( 
                 \frac{149753}{6} + 3564\zeta(3)
  \displaystyle 
 -\left[\frac{1078361}{162} + \frac{6508}{27}\zeta(3)\right]n_f
\right.
\\[3ex]
& \left.  \displaystyle 
 +\left[\frac{50065}{162}+ \frac{6472}{81}\zeta(3)\right]n_f^2 
 + \frac{1093}{729}n_f^3
                  \right)
{}\, .
\end{array} 
\end{equation} 

Eq.~(\ref{def:c-functions}) directly leads to the RG invariant $\mu$
independent mass $\hat{m}_q$
\begin{equation}
\hat{m}_q = \frac{m_q(\mu) }{c(a_s(\mu))}
\label{def1:RG-mass}
{}.
\end{equation}
An important property of $\hat{m}_q$ is its $\mu$ and scheme
independence. The latter follows from the fact that $\hat{m}_q$ could
be alternatively defined as follows
\begin{equation}
\hat{m}_q = \lim_{\mu \to \infty} 
m_q(\mu) \left( 
\frac{\alpha_s(\mu )}{\pi}\right )^{-\frac{\gamma_m^{0}}{\beta_0}}
\label{def2:RG-mass}
{}
\end{equation}
and from the well-known universality of the one loop coefficients of
the quark mass anomalous dimension and the $\beta$-function.

Evaluating the four loop approximation of the $c$-function in the $
\mathrm{RI} $ and $ \mathrm{RI'} $ schemes, we can state our results
for the conversion functions as a relation between the RG invariant
mass $\hat{m}$ and the masses $m^{ \mathrm{RI} }$ and $m^{
\mathrm{RI'} }$. For $n_f=3,4,5$ Eq.~(\ref{def1:RG-mass}) assumes the
form:

\begin{eqnarray}
\frac{\hat{m}^{(3)}}{m^{ \mathrm{RI} }} = \left(\frac{\alpha_s}{\pi}\right)^{-\frac{4}{9}}\Bigg\{ & & 
1 +  \frac{\alpha_s}{4\pi}
\left[ 
-\frac{722}{81}\right] 
 + \left(\frac{\alpha_s}{4\pi}\right)^{2}
\left[
-\frac{2521517}{13122} 
+\frac{536}{9}  \,\zeta_{3}
\right]
\nonumber\\
 &{+}& \left(\frac{\alpha_s}{4\pi}\right)^{3}
\left[
-\frac{88484924345}{12754584} 
  \right. \nonumber \\ &{}& \left.  
\phantom{+\left(\frac{\alpha_s}{4\pi}\right)^{3}}
+\frac{3089567}{972}  \,\zeta_{3}
-\frac{18640}{81}  \,\zeta_{5}
\right]
\,\Bigg\}{}
\label{MRGI3}
\end{eqnarray}

\begin{eqnarray}
\frac{\hat{m}^{(4)}}{m^{ \mathrm{RI} }} = \left(\frac{\alpha_s}{\pi}\right)^{-\frac{12}{25}}\Bigg\{ & & 
1 +  \frac{\alpha_s}{4\pi}
\left[ 
-\frac{17606}{1875}\right] 
\nonumber\\
 &{+}& \left(\frac{\alpha_s}{4\pi}\right)^{2}
\left[
-\frac{3819632767}{21093750} 
+\frac{952}{15}  \,\zeta_{3}
\right]
\nonumber\\
 &{+}& \left(\frac{\alpha_s}{4\pi}\right)^{3}
\left[
-\frac{8512503162869851}{1423828125000} 
  \right. \nonumber \\ &{}& \left.  
\phantom{+\left(\frac{\alpha_s}{4\pi}\right)^{3}}
+\frac{1035345331}{337500}  \,\zeta_{3}
-304  \,\zeta_{5}
\right]
\,\Bigg\}
\label{MRGI4}
\end{eqnarray}

\begin{eqnarray}
\frac{\hat{m}^{(5)}}{m^{ \mathrm{RI} }} = \left(\frac{\alpha_s}{\pi}\right)^{-\frac{12}{23}}\Bigg\{ & & 
1 +  \frac{\alpha_s}{4\pi}
\left[ 
-\frac{15926}{1587}\right] 
\nonumber\\
 &{+}& \left(\frac{\alpha_s}{4\pi}\right)^{2}
\left[
-\frac{2559841211}{15111414} 
+\frac{4696}{69}  \,\zeta_{3}
\right]
\nonumber\\
 &{+}& \left(\frac{\alpha_s}{4\pi}\right)^{3}
\left[
-\frac{4334826270205387}{863345304648} 
  \right. \nonumber \\ &{}& \left.  
\phantom{+\left(\frac{\alpha_s}{4\pi}\right)^{3}}
+\frac{3889063057}{1314036}  \,\zeta_{3}
-\frac{26960}{69}  \,\zeta_{5}
\right]
\,\Bigg\}{}.
\label{MRGI5}
\end{eqnarray}

\section{Conclusions}\label{sec:conclusion}

In this paper we have analytically computed the first few terms of the
high-energy expansion of the three-loop quark propagator.  These
results have been used to find the NNNLO conversion factors
transforming the $ \overline{\mathrm{MS}} $ quark mass and the
renormalized quark field to those defined in the $ \mathrm{RI} $
scheme which is more suitable for lattice QCD calculations.  The newly
computed NNNLO corrections are numerically significant and should be
taken into account when transforming the $ \mathrm{RI} $ quark masses
to the $ \overline{\mathrm{MS}} $ ones.

We also have presented the four loop results for the quark mass and
wave function anomalous dimensions in the $ \mathrm{RI} $ and $
\mathrm{RI'} $ schemes.  Unlike the case of
$\overline{\mathrm{MS}}$-scheme, the results display a striking
absence of the $\zeta_4$ irrational constant even at four loops.  This
could be attributed to the fact that the $RI/RI'$ quark mass and field
anomalous dimensions could be defined in a scheme-invariant way (see
subsection (\ref{sub:anomaldims4l}) for more details).
 
In principle, the knowledge of N${}^4$LO conversion factors would be
useful to even better control the convergence of the perturbation
series.  Unfortunately, such a calculation requires the knowledge of
the quark propagator at four loops --- a problem certainly out of the
range of present calculational techniques.

\section{Acknowledgments}
We would like to thank Oleg Veretin for his valuable comments on the
work \cite{Fleischer:1998dw} as well as for his help in the numerical
evaluation of the two loop fermion propagator.  One of us (K.G.Ch.) is
grateful to Vladimir Smirnov and Damir Becirevic for discussions and
advice. A.R. would like to thank Robert Harlander and Thorsten
Seidensticker for help and advice. We thank David Broadhurst and the
referee who both have drawn our attention to the absence of $\zeta_4$
in the quark mass and field anomalous dimensions in the RI and RI'
schemes.  This work was supported by DFG under contract Ku 502/8-1
({\it DFG-Forschergruppe ``Quantenfeldtheorie, Computeralgebra und
Monte-Carlo-Simulationen''}) and the {\it Gra\-duier\-ten\-kolleg
``Elementarteilchenphysik an Beschleunigern''}.

\section{Note}
In \cite{Becirevic:1999kb} the NNNLO $ \overline{\mathrm{MS}} $ --- $
\mathrm{RI} $ conversion relations have been used to transform the
lattice results for the $ \mathrm{RI} $ light quark masses into those
for the $ \overline{\mathrm{MS}} $ ones. In \cite{Broadhurst:1999zi}
the results for the (QED) fermion mass and field anomalous dimensions
( Eqs.~(\ref{FieldAnomalDimMS1qed},\ref{FieldAnomalDimMS2qed},
\ref{FieldAnomalDimMS3qed}) and Eqs.~(7-9) of \cite{Chetyrkin:1997dh})
have been reproduced within an entirely different approach.

\newpage

\begin{appendix}

\setcounter{equation}{0}

\section{Quark Propagators}\label{app:quarkprop3l}

Below we list the full three loop results for the quark
propagator\footnote{for the exact definition of the $\Sigma$'s see
Eqs.~(\ref{def:sigmas}) and (\ref{def:sigma-expansion})} computed in
general covariant gauge with the tree gluon propagator
\[
\frac{1}{q^2}(g_{{\mu\nu}}  - (1-\xi_L)q_\mu q_\nu/q^2)
{}.
\] 

For $ \mathrm{SU} (N)$ gauge group colour factors have the values $C_A
= N$, $C_F = (N^2-1)/(2N)$ and T = 1/2. The QED case is obtained with
substitutions $C_A \to 0$, $C_F \to 1$ and $T \to 1$.

\begin{eqnarray}\Sigma_S^{(1)} = & &  \,C_F 
\left[
-1 
-\frac{1}{2}  \,\xi_{L} 
+\frac{3}{4} l_{q \mu}
+\frac{1}{4}  \,\xi_{L} l_{q \mu}
\right]
{},
\label{SigmaS1}
\end{eqnarray}

\begin{eqnarray}\Sigma_S^{(2)} = & &  C_F^{2}
\left[
-\frac{13}{16} 
-\frac{3}{4}  \,\zeta_{3}
-\frac{1}{2}  \,\xi_{L} 
-\frac{1}{16}  \xi_{L}^{2}
+\frac{3}{4} l_{q \mu}
+\frac{5}{8}  \,\xi_{L} l_{q \mu}
+\frac{1}{8}  \xi_{L}^{2}l_{q \mu}
  \right. \nonumber \\ &{}& \left.  
\phantom{+ C_F^{2}}
-\frac{9}{32} l_{q \mu}^{2}
-\frac{3}{16}  \,\xi_{L} l_{q \mu}^{2}
-\frac{1}{32}  \xi_{L}^{2}l_{q \mu}^{2}
\right]
\nonumber\\
 &{+}&  \,C_F \,C_A 
\left[
-\frac{1531}{384} 
+\frac{21}{16}  \,\zeta_{3}
-\frac{5}{8}  \,\xi_{L} 
+\frac{3}{16}  \,\zeta_{3} \,\xi_{L} 
-\frac{15}{128}  \xi_{L}^{2}
+\frac{445}{192} l_{q \mu}
  \right. \nonumber \\ &{}& \left.  
\phantom{+ \,C_F \,C_A }
+\frac{5}{16}  \,\xi_{L} l_{q \mu}
+\frac{5}{64}  \xi_{L}^{2}l_{q \mu}
-\frac{11}{32} l_{q \mu}^{2}
-\frac{3}{64}  \,\xi_{L} l_{q \mu}^{2}
-\frac{1}{64}  \xi_{L}^{2}l_{q \mu}^{2}
\right]
\nonumber\\
 &{+}&  \,C_F \,T \,n_f 
\left[
\frac{13}{12} 
-\frac{2}{3} l_{q \mu}
+\frac{1}{8} l_{q \mu}^{2}
\right]
{},
\label{SigmaS2}
\end{eqnarray}

\begin{eqnarray}\Sigma_S^{(3)} = & &  C_F^{3}
\left[
-\frac{229}{48} 
-\frac{19}{32}  \,\zeta_{3}
+\frac{15}{8}  \,\zeta_{5}
-\frac{29}{64}  \,\xi_{L} 
-\frac{21}{32}  \,\zeta_{3} \,\xi_{L} 
-\frac{1}{16}  \xi_{L}^{2}
+\frac{3}{32}  \,\zeta_{3} \xi_{L}^{2}
  \right. \nonumber \\ &{}& \left.  
\phantom{+ C_F^{3}}
+\frac{1}{96}  \,\zeta_{3} \xi_{L}^{3}
+\frac{105}{64} l_{q \mu}
+\frac{9}{16}  \,\zeta_{3}l_{q \mu}
+\frac{37}{64}  \,\xi_{L} l_{q \mu}
+\frac{3}{16}  \,\zeta_{3} \,\xi_{L} l_{q \mu}
  \right. \nonumber \\ &{}& \left.  
\phantom{+ C_F^{3}}
+\frac{11}{64}  \xi_{L}^{2}l_{q \mu}
+\frac{1}{64}  \xi_{L}^{3}l_{q \mu}
-\frac{9}{32} l_{q \mu}^{2}
-\frac{21}{64}  \,\xi_{L} l_{q \mu}^{2}
-\frac{1}{8}  \xi_{L}^{2}l_{q \mu}^{2}
  \right. \nonumber \\ &{}& \left.  
\phantom{+ C_F^{3}}
-\frac{1}{64}  \xi_{L}^{3}l_{q \mu}^{2}
+\frac{9}{128} l_{q \mu}^{3}
+\frac{9}{128}  \,\xi_{L} l_{q \mu}^{3}
+\frac{3}{128}  \xi_{L}^{2}l_{q \mu}^{3}
+\frac{1}{384}  \xi_{L}^{3}l_{q \mu}^{3}
\right]
\nonumber\\
 &{+}&  C_F^{2}\,C_A 
\left[
-\frac{3005}{1152} 
-\frac{313}{96}  \,\zeta_{3}
-\frac{3}{32}  \,\zeta_{4}
+\frac{5}{8}  \,\zeta_{5}
-\frac{1861}{768}  \,\xi_{L} 
+\frac{43}{64}  \,\zeta_{3} \,\xi_{L} 
  \right. \nonumber \\ &{}& \left.  
\phantom{+ C_F^{2}\,C_A }
-\frac{5}{16}  \,\zeta_{5} \,\xi_{L} 
-\frac{35}{128}  \xi_{L}^{2}
-\frac{5}{256}  \xi_{L}^{3}
-\frac{1}{64}  \,\zeta_{3} \xi_{L}^{3}
+\frac{2467}{512} l_{q \mu}
  \right. \nonumber \\ &{}& \left.  
\phantom{+ C_F^{2}\,C_A }
+\frac{37}{64}  \,\zeta_{3}l_{q \mu}
+\frac{4511}{1536}  \,\xi_{L} l_{q \mu}
-\frac{15}{32}  \,\zeta_{3} \,\xi_{L} l_{q \mu}
+\frac{221}{512}  \xi_{L}^{2}l_{q \mu}
  \right. \nonumber \\ &{}& \left.  
\phantom{+ C_F^{2}\,C_A }
-\frac{3}{64}  \,\zeta_{3} \xi_{L}^{2}l_{q \mu}
+\frac{27}{512}  \xi_{L}^{3}l_{q \mu}
-\frac{533}{256} l_{q \mu}^{2}
-\frac{793}{768}  \,\xi_{L} l_{q \mu}^{2}
  \right. \nonumber \\ &{}& \left.  
\phantom{+ C_F^{2}\,C_A }
-\frac{45}{256}  \xi_{L}^{2}l_{q \mu}^{2}
-\frac{7}{256}  \xi_{L}^{3}l_{q \mu}^{2}
+\frac{33}{128} l_{q \mu}^{3}
+\frac{31}{256}  \,\xi_{L} l_{q \mu}^{3}
  \right. \nonumber \\ &{}& \left.  
\phantom{+ C_F^{2}\,C_A }
+\frac{3}{128}  \xi_{L}^{2}l_{q \mu}^{3}
+\frac{1}{256}  \xi_{L}^{3}l_{q \mu}^{3}
\right]
\nonumber\\
 &{+}&  C_F^{2}\,T \,n_f 
\left[
\frac{4699}{1152} 
-\frac{11}{12}  \,\zeta_{3}
-\frac{3}{8}  \,\zeta_{4}
+\frac{23}{48}  \,\xi_{L} 
+\frac{1}{8}  \,\zeta_{3} \,\xi_{L} 
-\frac{167}{64} l_{q \mu}
  \right. \nonumber \\ &{}& \left.  
\phantom{+ C_F^{2}\,T \,n_f }
+\frac{1}{4}  \,\zeta_{3}l_{q \mu}
-\frac{29}{48}  \,\xi_{L} l_{q \mu}
+\frac{23}{32} l_{q \mu}^{2}
+\frac{11}{48}  \,\xi_{L} l_{q \mu}^{2}
-\frac{3}{32} l_{q \mu}^{3}
  \right. \nonumber \\ &{}& \left.  
\phantom{+ C_F^{2}\,T \,n_f }
-\frac{1}{32}  \,\xi_{L} l_{q \mu}^{3}
\right]
\nonumber\\
 &{+}&  \,C_F C_A^{2}
\left[
-\frac{4315565}{248832} 
+\frac{20305}{2304}  \,\zeta_{3}
+\frac{69}{1024}  \,\zeta_{4}
-\frac{405}{256}  \,\zeta_{5}
-\frac{66301}{36864}  \,\xi_{L} 
  \right. \nonumber \\ &{}& \left.  
\phantom{+ \,C_F C_A^{2}}
+\frac{193}{256}  \,\zeta_{3} \,\xi_{L} 
-\frac{3}{512}  \,\zeta_{4} \,\xi_{L} 
-\frac{5}{128}  \,\zeta_{5} \,\xi_{L} 
-\frac{1377}{4096}  \xi_{L}^{2}
  \right. \nonumber \\ &{}& \left.  
\phantom{+ \,C_F C_A^{2}}
+\frac{3}{32}  \,\zeta_{3} \xi_{L}^{2}
-\frac{3}{1024}  \,\zeta_{4} \xi_{L}^{2}
-\frac{5}{256}  \,\zeta_{5} \xi_{L}^{2}
-\frac{191}{3072}  \xi_{L}^{3}
  \right. \nonumber \\ &{}& \left.  
\phantom{+ \,C_F C_A^{2}}
+\frac{1}{192}  \,\zeta_{3} \xi_{L}^{3}
+\frac{294793}{27648} l_{q \mu}
-\frac{1301}{512}  \,\zeta_{3}l_{q \mu}
+\frac{6509}{6144}  \,\xi_{L} l_{q \mu}
  \right. \nonumber \\ &{}& \left.  
\phantom{+ \,C_F C_A^{2}}
-\frac{59}{256}  \,\zeta_{3} \,\xi_{L} l_{q \mu}
+\frac{467}{2048}  \xi_{L}^{2}l_{q \mu}
-\frac{9}{512}  \,\zeta_{3} \xi_{L}^{2}l_{q \mu}
+\frac{43}{1024}  \xi_{L}^{3}l_{q \mu}
  \right. \nonumber \\ &{}& \left.  
\phantom{+ \,C_F C_A^{2}}
-\frac{5507}{2304} l_{q \mu}^{2}
-\frac{715}{3072}  \,\xi_{L} l_{q \mu}^{2}
-\frac{61}{1024}  \xi_{L}^{2}l_{q \mu}^{2}
-\frac{3}{256}  \xi_{L}^{3}l_{q \mu}^{2}
  \right. \nonumber \\ &{}& \left.  
\phantom{+ \,C_F C_A^{2}}
+\frac{121}{576} l_{q \mu}^{3}
+\frac{31}{1536}  \,\xi_{L} l_{q \mu}^{3}
+\frac{3}{512}  \xi_{L}^{2}l_{q \mu}^{3}
+\frac{1}{768}  \xi_{L}^{3}l_{q \mu}^{3}
\right]
\nonumber\\
 &{+}&  \,C_F \,C_A \,T \,n_f 
\left[
\frac{16381}{1944} 
-\frac{193}{144}  \,\zeta_{3}
+\frac{3}{8}  \,\zeta_{4}
+\frac{2773}{4608}  \,\xi_{L} 
-\frac{3}{16}  \,\zeta_{3} \,\xi_{L} 
  \right. \nonumber \\ &{}& \left.  
\phantom{+ \,C_F \,C_A \,T \,n_f }
-\frac{5081}{864} l_{q \mu}
+\frac{1}{8}  \,\zeta_{3}l_{q \mu}
-\frac{251}{768}  \,\xi_{L} l_{q \mu}
+\frac{1}{16}  \,\zeta_{3} \,\xi_{L} l_{q \mu}
  \right. \nonumber \\ &{}& \left.  
\phantom{+ \,C_F \,C_A \,T \,n_f }
+\frac{887}{576} l_{q \mu}^{2}
+\frac{25}{384}  \,\xi_{L} l_{q \mu}^{2}
-\frac{11}{72} l_{q \mu}^{3}
-\frac{1}{192}  \,\xi_{L} l_{q \mu}^{3}
\right]
\nonumber\\
 &{+}&  \,C_F  T^{2} \, n_f^{2}
\left[
-\frac{191}{243} 
-\frac{1}{18}  \,\zeta_{3}
+\frac{73}{108} l_{q \mu}
-\frac{2}{9} l_{q \mu}^{2}
+\frac{1}{36} l_{q \mu}^{3}
\right]
{},
\label{SigmaS3}
\end{eqnarray}

\begin{eqnarray}\Sigma_V^{(1)} = & &  \,C_F 
\left[
\frac{1}{4}  \,\xi_{L} 
-\frac{1}{4}  \,\xi_{L} l_{q \mu}
\right]
{},
\label{SigmaV1}
\end{eqnarray}

\begin{eqnarray}\Sigma_V^{(2)} = & &  C_F^{2}
\left[
-\frac{5}{128} 
+\frac{3}{32} l_{q \mu}
-\frac{1}{16}  \xi_{L}^{2}l_{q \mu}
+\frac{1}{32}  \xi_{L}^{2}l_{q \mu}^{2}
\right]
\nonumber\\
 &{+}&  \,C_F \,C_A 
\left[
\frac{41}{64} 
-\frac{3}{16}  \,\zeta_{3}
+\frac{13}{32}  \,\xi_{L} 
-\frac{3}{16}  \,\zeta_{3} \,\xi_{L} 
+\frac{9}{128}  \xi_{L}^{2}
-\frac{25}{64} l_{q \mu}
  \right. \nonumber \\ &{}& \left.  
\phantom{+ \,C_F \,C_A }
-\frac{7}{32}  \,\xi_{L} l_{q \mu}
-\frac{3}{64}  \xi_{L}^{2}l_{q \mu}
+\frac{3}{64}  \,\xi_{L} l_{q \mu}^{2}
+\frac{1}{64}  \xi_{L}^{2}l_{q \mu}^{2}
\right]
\nonumber\\
 &{+}&  \,C_F \,T \,n_f 
\left[
-\frac{7}{32} 
+\frac{1}{8} l_{q \mu}
\right]
{},
\label{SigmaV2}
\end{eqnarray}

\begin{eqnarray}\Sigma_V^{(3)} = & &  C_F^{3}
\left[
-\frac{73}{768} 
+\frac{7}{512}  \,\xi_{L} 
-\frac{1}{96}  \,\zeta_{3} \xi_{L}^{3}
-\frac{3}{128} l_{q \mu}
+\frac{17}{512}  \,\xi_{L} l_{q \mu}
  \right. \nonumber \\ &{}& \left.  
\phantom{+ C_F^{3}}
-\frac{3}{128}  \,\xi_{L} l_{q \mu}^{2}
+\frac{1}{128}  \xi_{L}^{3}l_{q \mu}^{2}
-\frac{1}{384}  \xi_{L}^{3}l_{q \mu}^{3}
\right]
\nonumber\\
 &{+}&  C_F^{2}\,C_A 
\left[
-\frac{997}{1536} 
+\frac{11}{16}  \,\zeta_{3}
+\frac{3}{32}  \,\zeta_{4}
-\frac{5}{16}  \,\zeta_{5}
+\frac{1}{16}  \,\xi_{L} 
-\frac{17}{64}  \,\zeta_{3} \,\xi_{L} 
  \right. \nonumber \\ &{}& \left.  
\phantom{+ C_F^{2}\,C_A }
+\frac{5}{16}  \,\zeta_{5} \,\xi_{L} 
+\frac{3}{128}  \xi_{L}^{2}
-\frac{1}{512}  \xi_{L}^{3}
+\frac{1}{64}  \,\zeta_{3} \xi_{L}^{3}
+\frac{121}{192} l_{q \mu}
  \right. \nonumber \\ &{}& \left.  
\phantom{+ C_F^{2}\,C_A }
-\frac{3}{16}  \,\zeta_{3}l_{q \mu}
-\frac{33}{128}  \,\xi_{L} l_{q \mu}
+\frac{3}{64}  \,\zeta_{3} \,\xi_{L} l_{q \mu}
-\frac{17}{128}  \xi_{L}^{2}l_{q \mu}
  \right. \nonumber \\ &{}& \left.  
\phantom{+ C_F^{2}\,C_A }
+\frac{3}{64}  \,\zeta_{3} \xi_{L}^{2}l_{q \mu}
-\frac{11}{512}  \xi_{L}^{3}l_{q \mu}
-\frac{11}{128} l_{q \mu}^{2}
+\frac{25}{256}  \,\xi_{L} l_{q \mu}^{2}
  \right. \nonumber \\ &{}& \left.  
\phantom{+ C_F^{2}\,C_A }
+\frac{17}{256}  \xi_{L}^{2}l_{q \mu}^{2}
+\frac{1}{64}  \xi_{L}^{3}l_{q \mu}^{2}
-\frac{3}{256}  \xi_{L}^{2}l_{q \mu}^{3}
-\frac{1}{256}  \xi_{L}^{3}l_{q \mu}^{3}
\right]
\nonumber\\
 &{+}&  C_F^{2}\,T \,n_f 
\left[
-\frac{79}{384} 
+\frac{1}{4}  \,\zeta_{3}
-\frac{3}{128}  \,\xi_{L} 
-\frac{7}{96} l_{q \mu}
+\frac{11}{128}  \,\xi_{L} l_{q \mu}
  \right. \nonumber \\ &{}& \left.  
\phantom{+ C_F^{2}\,T \,n_f }
+\frac{1}{32} l_{q \mu}^{2}
-\frac{1}{32}  \,\xi_{L} l_{q \mu}^{2}
\right]
\nonumber\\
 &{+}&  \,C_F C_A^{2}
\left[
\frac{159257}{41472} 
-\frac{3139}{1536}  \,\zeta_{3}
-\frac{69}{1024}  \,\zeta_{4}
+\frac{165}{256}  \,\zeta_{5}
+\frac{39799}{36864}  \,\xi_{L} 
  \right. \nonumber \\ &{}& \left.  
\phantom{+ \,C_F C_A^{2}}
-\frac{35}{64}  \,\zeta_{3} \,\xi_{L} 
+\frac{3}{512}  \,\zeta_{4} \,\xi_{L} 
+\frac{5}{128}  \,\zeta_{5} \,\xi_{L} 
+\frac{787}{4096}  \xi_{L}^{2}
  \right. \nonumber \\ &{}& \left.  
\phantom{+ \,C_F C_A^{2}}
-\frac{39}{512}  \,\zeta_{3} \xi_{L}^{2}
+\frac{3}{1024}  \,\zeta_{4} \xi_{L}^{2}
+\frac{5}{256}  \,\zeta_{5} \xi_{L}^{2}
+\frac{55}{1536}  \xi_{L}^{3}
  \right. \nonumber \\ &{}& \left.  
\phantom{+ \,C_F C_A^{2}}
-\frac{1}{192}  \,\zeta_{3} \xi_{L}^{3}
-\frac{19979}{9216} l_{q \mu}
+\frac{245}{512}  \,\zeta_{3}l_{q \mu}
-\frac{4393}{6144}  \,\xi_{L} l_{q \mu}
  \right. \nonumber \\ &{}& \left.  
\phantom{+ \,C_F C_A^{2}}
+\frac{59}{256}  \,\zeta_{3} \,\xi_{L} l_{q \mu}
-\frac{295}{2048}  \xi_{L}^{2}l_{q \mu}
+\frac{9}{512}  \,\zeta_{3} \xi_{L}^{2}l_{q \mu}
-\frac{27}{1024}  \xi_{L}^{3}l_{q \mu}
  \right. \nonumber \\ &{}& \left.  
\phantom{+ \,C_F C_A^{2}}
+\frac{275}{768} l_{q \mu}^{2}
+\frac{529}{3072}  \,\xi_{L} l_{q \mu}^{2}
+\frac{43}{1024}  \xi_{L}^{2}l_{q \mu}^{2}
+\frac{1}{128}  \xi_{L}^{3}l_{q \mu}^{2}
  \right. \nonumber \\ &{}& \left.  
\phantom{+ \,C_F C_A^{2}}
-\frac{31}{1536}  \,\xi_{L} l_{q \mu}^{3}
-\frac{3}{512}  \xi_{L}^{2}l_{q \mu}^{3}
-\frac{1}{768}  \xi_{L}^{3}l_{q \mu}^{3}
\right]
\nonumber\\
 &{+}&  \,C_F \,C_A \,T \,n_f 
\left[
-\frac{11887}{5184} 
+\frac{13}{48}  \,\zeta_{3}
-\frac{1723}{4608}  \,\xi_{L} 
+\frac{1}{8}  \,\zeta_{3} \,\xi_{L} 
  \right. \nonumber \\ &{}& \left.  
\phantom{+ \,C_F \,C_A \,T \,n_f }
+\frac{191}{144} l_{q \mu}
-\frac{1}{8}  \,\zeta_{3}l_{q \mu}
+\frac{175}{768}  \,\xi_{L} l_{q \mu}
-\frac{1}{16}  \,\zeta_{3} \,\xi_{L} l_{q \mu}
  \right. \nonumber \\ &{}& \left.  
\phantom{+ \,C_F \,C_A \,T \,n_f }
-\frac{47}{192} l_{q \mu}^{2}
-\frac{19}{384}  \,\xi_{L} l_{q \mu}^{2}
+\frac{1}{192}  \,\xi_{L} l_{q \mu}^{3}
\right]
\nonumber\\
 &{+}&  \,C_F  T^{2} \, n_f^{2}
\left[
\frac{785}{2592} 
-\frac{13}{72} l_{q \mu}
+\frac{1}{24} l_{q \mu}^{2}
\right]
{}.
\label{SigmaV3}
\end{eqnarray}

\setcounter{equation}{0}

\section{Conversion Functions}\label{app:conversion}

The full gauge dependent conversion factors read\footnote{Note that
these results are expanded in $a_s = \frac{\alpha_s}{\pi}$ and not in
$\frac{\alpha_s}{4 \pi}$ as in Eqs.~(\ref{C2RIqcd}) to
(\ref{CmRIpqcd}).}

\begin{eqnarray}C_2^{ \mathrm{RI}  (1)} = & &  \,C_F 
\left[
-\frac{1}{8}  \,\xi_{L} 
\right]
{},
\label{C2RI1}
\end{eqnarray}

\begin{eqnarray}C_2^{ \mathrm{RI}  (2)} = & & \,C_A  \,C_F 
\left[
-\frac{57}{128} 
+\frac{3}{16}  \,\zeta_{3}
-\frac{19}{64}  \,\xi_{L} 
+\frac{3}{16}  \,\zeta_{3} \,\xi_{L} 
-\frac{3}{64}  \xi_{L}^{2}
\right]
\nonumber\\
 &{+}&  C_F^{2}
\left[
-\frac{1}{128} 
+\frac{3}{64}  \xi_{L}^{2}
\right]
 \, + \,  \,C_F  \,n_f \,T
\left[ 
 \frac{5}{32}\right] 
{},
\label{C2RI2}
\end{eqnarray}

\begin{eqnarray}C_2^{ \mathrm{RI}  (3)} = & & C_A^{2} \,C_F 
\left[
-\frac{457217}{165888} 
+\frac{5543}{3072}  \,\zeta_{3}
+\frac{69}{1024}  \,\zeta_{4}
-\frac{165}{256}  \,\zeta_{5}
-\frac{6655}{9216}  \,\xi_{L} 
  \right. \nonumber \\ &{}& \left.  
\phantom{+C_A^{2} \,C_F }
+\frac{221}{512}  \,\zeta_{3} \,\xi_{L} 
-\frac{3}{512}  \,\zeta_{4} \,\xi_{L} 
-\frac{5}{128}  \,\zeta_{5} \,\xi_{L} 
-\frac{123}{1024}  \xi_{L}^{2}
  \right. \nonumber \\ &{}& \left.  
\phantom{+C_A^{2} \,C_F }
+\frac{69}{1024}  \,\zeta_{3} \xi_{L}^{2}
-\frac{3}{1024}  \,\zeta_{4} \xi_{L}^{2}
-\frac{5}{256}  \,\zeta_{5} \xi_{L}^{2}
-\frac{139}{6144}  \xi_{L}^{3}
  \right. \nonumber \\ &{}& \left.  
\phantom{+C_A^{2} \,C_F }
+\frac{1}{192}  \,\zeta_{3} \xi_{L}^{3}
\right]
\nonumber\\
 &{+}& \,C_A  C_F^{2}
\left[
\frac{171}{512} 
-\frac{19}{32}  \,\zeta_{3}
-\frac{3}{32}  \,\zeta_{4}
+\frac{5}{16}  \,\zeta_{5}
+\frac{91}{512}  \,\xi_{L} 
+\frac{25}{128}  \,\zeta_{3} \,\xi_{L} 
  \right. \nonumber \\ &{}& \left.  
\phantom{+\,C_A  C_F^{2}}
-\frac{5}{16}  \,\zeta_{5} \,\xi_{L} 
+\frac{15}{128}  \xi_{L}^{2}
-\frac{9}{128}  \,\zeta_{3} \xi_{L}^{2}
+\frac{25}{1024}  \xi_{L}^{3}
  \right. \nonumber \\ &{}& \left.  
\phantom{+\,C_A  C_F^{2}}
-\frac{1}{64}  \,\zeta_{3} \xi_{L}^{3}
\right]
\nonumber\\
 &{+}&  C_F^{3}
\left[
\frac{41}{384} 
-\frac{29}{1024}  \,\xi_{L} 
-\frac{5}{512}  \xi_{L}^{3}
+\frac{1}{96}  \,\zeta_{3} \xi_{L}^{3}
\right]
\nonumber\\
 &{+}& \,C_A  \,C_F  \,n_f \,T
\left[
\frac{8449}{5184} 
-\frac{5}{24}  \,\zeta_{3}
+\frac{599}{2304}  \,\xi_{L} 
-\frac{3}{32}  \,\zeta_{3} \,\xi_{L} 
\right]
\nonumber\\
 &{+}&  C_F^{2} \,n_f \,T
\left[
\frac{31}{128} 
-\frac{1}{4}  \,\zeta_{3}
-\frac{15}{256}  \,\xi_{L} 
\right]
 \, + \,  \,C_F  \, n_f^{2} T^{2}
\left[ 
-\frac{551}{2592}\right] 
{},
\label{C2RI3}
\end{eqnarray}

\begin{eqnarray}C_m^{ \mathrm{RI}  (1)} = & &  \,C_F 
\left[
-1 
-\frac{3}{8}  \,\xi_{L} 
\right]
{},
\label{CmRI1}
\end{eqnarray}

\begin{eqnarray}C_m^{ \mathrm{RI}  (2)} = & & \,C_A  \,C_F 
\left[
-\frac{85}{24} 
+\frac{9}{8}  \,\zeta_{3}
-\frac{21}{64}  \,\xi_{L} 
-\frac{9}{128}  \xi_{L}^{2}
\right]
\nonumber\\
 &{+}&  C_F^{2}
\left[
\frac{25}{128} 
-\frac{3}{4}  \,\zeta_{3}
+\frac{3}{8}  \,\xi_{L} 
+\frac{3}{32}  \xi_{L}^{2}
\right]
 \, + \,  \,C_F  \,n_f \,T
\left[ 
 \frac{89}{96}\right] 
{},
\label{CmRI2}
\end{eqnarray}

\begin{eqnarray}C_m^{ \mathrm{RI}  (3)} = & & C_A^{2} \,C_F 
\left[
-\frac{7259479}{497664} 
+\frac{64591}{9216}  \,\zeta_{3}
-\frac{15}{16}  \,\zeta_{5}
-\frac{4409}{4096}  \,\xi_{L} 
  \right. \nonumber \\ &{}& \left.  
\phantom{+C_A^{2} \,C_F }
+\frac{165}{512}  \,\zeta_{3} \,\xi_{L} 
-\frac{885}{4096}  \xi_{L}^{2}
+\frac{27}{1024}  \,\zeta_{3} \xi_{L}^{2}
-\frac{81}{2048}  \xi_{L}^{3}
\right]
\nonumber\\
 &{+}& \,C_A  C_F^{2}
\left[
\frac{21133}{4608} 
-\frac{245}{48}  \,\zeta_{3}
+\frac{5}{16}  \,\zeta_{5}
+\frac{1771}{1024}  \,\xi_{L} 
-\frac{13}{16}  \,\zeta_{3} \,\xi_{L} 
  \right. \nonumber \\ &{}& \left.  
\phantom{+\,C_A  C_F^{2}}
+\frac{69}{256}  \xi_{L}^{2}
-\frac{3}{64}  \,\zeta_{3} \xi_{L}^{2}
+\frac{3}{64}  \xi_{L}^{3}
\right]
\nonumber\\
 &{+}&  C_F^{3}
\left[
-\frac{409}{96} 
+\frac{29}{32}  \,\zeta_{3}
+\frac{15}{8}  \,\zeta_{5}
-\frac{93}{1024}  \,\xi_{L} 
-\frac{3}{32}  \xi_{L}^{2}
+\frac{3}{32}  \,\zeta_{3} \xi_{L}^{2}
  \right. \nonumber \\ &{}& \left.  
\phantom{+ C_F^{3}}
-\frac{3}{128}  \xi_{L}^{3}
\right]
\nonumber\\
 &{+}& \,C_A  \,C_F  \,n_f \,T
\left[
\frac{105701}{15552} 
-\frac{163}{144}  \,\zeta_{3}
+\frac{3}{8}  \,\zeta_{4}
+\frac{175}{512}  \,\xi_{L} 
-\frac{3}{32}  \,\zeta_{3} \,\xi_{L} 
\right]
\nonumber\\
 &{+}&  C_F^{2} \,n_f \,T
\left[
\frac{263}{144} 
-\frac{2}{3}  \,\zeta_{3}
-\frac{3}{8}  \,\zeta_{4}
-\frac{95}{256}  \,\xi_{L} 
+\frac{1}{8}  \,\zeta_{3} \,\xi_{L} 
\right]
\nonumber\\
 &{+}&  \,C_F  \, n_f^{2} T^{2}
\left[
-\frac{4459}{7776} 
-\frac{1}{18}  \,\zeta_{3}
\right]
{},
\label{CmRI3}
\end{eqnarray}

\begin{eqnarray}C_2^{ \mathrm{RI'}  (1)} = & &  \,C_F 
\left[
-\frac{1}{4}  \,\xi_{L} 
\right]
{},
\label{C2RIp1}
\end{eqnarray}

\begin{eqnarray}C_2^{ \mathrm{RI'}  (2)} = & & \,C_A  \,C_F 
\left[
-\frac{41}{64} 
+\frac{3}{16}  \,\zeta_{3}
-\frac{13}{32}  \,\xi_{L} 
+\frac{3}{16}  \,\zeta_{3} \,\xi_{L} 
-\frac{9}{128}  \xi_{L}^{2}
\right]
\nonumber\\
 &{+}&  C_F^{2}
\left[
\frac{5}{128} 
+\frac{1}{16}  \xi_{L}^{2}
\right]
 \, + \,  \,C_F  \,n_f \,T
\left[ 
 \frac{7}{32}\right] 
{},
\label{C2RIp2}
\end{eqnarray}

\begin{eqnarray}C_2^{ \mathrm{RI'}  (3)} = & & \,C_A  C_F^{2}
\left[
\frac{997}{1536} 
-\frac{11}{16}  \,\zeta_{3}
-\frac{3}{32}  \,\zeta_{4}
+\frac{5}{16}  \,\zeta_{5}
+\frac{33}{128}  \,\xi_{L} 
+\frac{11}{64}  \,\zeta_{3} \,\xi_{L} 
  \right. \nonumber \\ &{}& \left.  
\phantom{+\,C_A  C_F^{2}}
-\frac{5}{16}  \,\zeta_{5} \,\xi_{L} 
+\frac{23}{128}  \xi_{L}^{2}
-\frac{3}{32}  \,\zeta_{3} \xi_{L}^{2}
+\frac{19}{512}  \xi_{L}^{3}
-\frac{1}{64}  \,\zeta_{3} \xi_{L}^{3}
\right]
\nonumber\\
 &{+}& C_A^{2} \,C_F 
\left[
-\frac{159257}{41472} 
+\frac{3139}{1536}  \,\zeta_{3}
+\frac{69}{1024}  \,\zeta_{4}
-\frac{165}{256}  \,\zeta_{5}
  \right. \nonumber \\ &{}& \left.  
\phantom{+C_A^{2} \,C_F }
-\frac{39799}{36864}  \,\xi_{L} 
+\frac{35}{64}  \,\zeta_{3} \,\xi_{L} 
-\frac{3}{512}  \,\zeta_{4} \,\xi_{L} 
-\frac{5}{128}  \,\zeta_{5} \,\xi_{L} 
  \right. \nonumber \\ &{}& \left.  
\phantom{+C_A^{2} \,C_F }
-\frac{787}{4096}  \xi_{L}^{2}
+\frac{39}{512}  \,\zeta_{3} \xi_{L}^{2}
-\frac{3}{1024}  \,\zeta_{4} \xi_{L}^{2}
-\frac{5}{256}  \,\zeta_{5} \xi_{L}^{2}
  \right. \nonumber \\ &{}& \left.  
\phantom{+C_A^{2} \,C_F }
-\frac{55}{1536}  \xi_{L}^{3}
+\frac{1}{192}  \,\zeta_{3} \xi_{L}^{3}
\right]
\nonumber\\
 &{+}&  C_F^{3}
\left[
\frac{73}{768} 
-\frac{17}{512}  \,\xi_{L} 
-\frac{1}{64}  \xi_{L}^{3}
+\frac{1}{96}  \,\zeta_{3} \xi_{L}^{3}
\right]
\nonumber\\
 &{+}& \,C_A  \,C_F  \,n_f \,T
\left[
\frac{11887}{5184} 
-\frac{13}{48}  \,\zeta_{3}
+\frac{1723}{4608}  \,\xi_{L} 
-\frac{1}{8}  \,\zeta_{3} \,\xi_{L} 
\right]
\nonumber\\
 &{+}&  C_F^{2} \,n_f \,T
\left[
\frac{79}{384} 
-\frac{1}{4}  \,\zeta_{3}
-\frac{11}{128}  \,\xi_{L} 
\right]
 \, + \,  \,C_F  \, n_f^{2} T^{2}
\left[ 
-\frac{785}{2592}\right] 
{},
\label{C2RIp3}
\end{eqnarray}

\begin{eqnarray}C_m^{ \mathrm{RI'}  (1)} = & &  \,C_F 
\left[
-1 
-\frac{1}{4}  \,\xi_{L} 
\right]
{},
\label{CmRIp1}
\end{eqnarray}

\begin{eqnarray}C_m^{ \mathrm{RI'}  (2)} = & & \,C_A  \,C_F 
\left[
-\frac{1285}{384} 
+\frac{9}{8}  \,\zeta_{3}
-\frac{7}{32}  \,\xi_{L} 
-\frac{3}{64}  \xi_{L}^{2}
\right]
\nonumber\\
 &{+}&  C_F^{2}
\left[
\frac{19}{128} 
-\frac{3}{4}  \,\zeta_{3}
+\frac{1}{4}  \,\xi_{L} 
+\frac{1}{16}  \xi_{L}^{2}
\right]
 \quad + \quad  \,C_F  \,n_f \,T
\left[ 
 \frac{83}{96}\right] 
{},
\label{CmRIp2}
\end{eqnarray}

\begin{eqnarray}C_m^{ \mathrm{RI'}  (3)} = & & C_A^{2} \,C_F 
\left[
-\frac{3360023}{248832} 
+\frac{31193}{4608}  \,\zeta_{3}
-\frac{15}{16}  \,\zeta_{5}
-\frac{4417}{6144}  \,\xi_{L} 
+\frac{53}{256}  \,\zeta_{3} \,\xi_{L} 
  \right. \nonumber \\ &{}& \left.  
\phantom{+C_A^{2} \,C_F }
-\frac{295}{2048}  \xi_{L}^{2}
+\frac{9}{512}  \,\zeta_{3} \xi_{L}^{2}
-\frac{27}{1024}  \xi_{L}^{3}
\right]
\nonumber\\
 &{+}& \,C_A  C_F^{2}
\left[
\frac{18781}{4608} 
-\frac{481}{96}  \,\zeta_{3}
+\frac{5}{16}  \,\zeta_{5}
+\frac{1771}{1536}  \,\xi_{L} 
-\frac{43}{64}  \,\zeta_{3} \,\xi_{L} 
  \right. \nonumber \\ &{}& \left.  
\phantom{+\,C_A  C_F^{2}}
+\frac{23}{128}  \xi_{L}^{2}
-\frac{3}{64}  \,\zeta_{3} \xi_{L}^{2}
+\frac{1}{32}  \xi_{L}^{3}
\right]
\nonumber\\
 &{+}&  C_F^{3}
\left[
-\frac{3227}{768} 
+\frac{29}{32}  \,\zeta_{3}
+\frac{15}{8}  \,\zeta_{5}
-\frac{31}{512}  \,\xi_{L} 
-\frac{3}{32}  \,\zeta_{3} \,\xi_{L} 
-\frac{1}{16}  \xi_{L}^{2}
  \right. \nonumber \\ &{}& \left.  
\phantom{+ C_F^{3}}
+\frac{3}{32}  \,\zeta_{3} \xi_{L}^{2}
-\frac{1}{64}  \xi_{L}^{3}
\right]
\nonumber\\
 &{+}& \,C_A  \,C_F  \,n_f \,T
\left[
\frac{95387}{15552} 
-\frac{77}{72}  \,\zeta_{3}
+\frac{3}{8}  \,\zeta_{4}
+\frac{175}{768}  \,\xi_{L} 
-\frac{1}{16}  \,\zeta_{3} \,\xi_{L} 
\right]
\nonumber\\
 &{+}&  C_F^{2} \,n_f \,T
\left[
\frac{1109}{576} 
-\frac{2}{3}  \,\zeta_{3}
-\frac{3}{8}  \,\zeta_{4}
-\frac{95}{384}  \,\xi_{L} 
+\frac{1}{8}  \,\zeta_{3} \,\xi_{L} 
\right]
\nonumber\\
 &{+}&  \,C_F  \, n_f^{2} T^{2}
\left[
-\frac{3757}{7776} 
-\frac{1}{18}  \,\zeta_{3}
\right]
{}.
\label{CmRIp3}
\end{eqnarray}

\setcounter{equation}{0}

\section{Anomalous Dimensions}\label{app:anomaldims4l}

The quark mass and field anomalous dimensions for general $
\mathrm{SU} (N)$ and for the $ \overline{\mathrm{MS}} $, $ \mathrm{RI}
$ and $ \mathrm{RI'} $ schemes are listed below. See
Eq.~(\ref{MassAnomalDimMS0qcd}) for the conventions. For the $
\overline{\mathrm{MS}} $ case also the field anomalous dimension
$\gamma_{2}$ is given for general gauge and $ \mathrm{SU} (N)$, while
for the $ \mathrm{RI} $ and $ \mathrm{RI'} $ only the Landau-gauge
formulae are given (for which all $\gamma_2^{(0)}$ are $0$):

\begin{eqnarray}\gamma_{m}^{(0)} = \frac{N^2 - 1}{N}\Bigg\{ & & 
 \frac{3}{8}\,\Bigg\}{},
\label{MassAnomalDimMS0}
\end{eqnarray}

\begin{eqnarray}\gamma_{m}^{(1)} = \frac{N^2-1}{16 N^2}\Bigg\{ & & 
\left[
-\frac{3}{8} 
+\frac{203}{24}  N^{2}
\right]
 \, + \,  \,n_f 
\left[
-\frac{5}{6}  N 
\right]
\,\Bigg\}{},
\label{MassAnomalDimMS1}
\end{eqnarray}

\begin{eqnarray}\gamma_{m}^{(2)} = \frac{N^2-1}{64 N^3}\Bigg\{ & & 
\left[
\frac{129}{16} 
-\frac{129}{16}  N^{2}
+\frac{11413}{216}  N^{4}
\right]
\nonumber\\
 &{+}&  \,n_f 
\left[
\frac{23}{4}  N 
-\frac{1177}{108}  N^{3}
-6  N  \,\zeta_{3}
-6  N^{3} \,\zeta_{3}
\right]
\nonumber\\
 &{+}&  \, n_f^{2}
\left[
-\frac{35}{54}  N^{2}
\right]
\,\Bigg\}{},
\label{MassAnomalDimMS2}
\end{eqnarray}

\begin{eqnarray}\gamma_{m}^{(3)} = \frac{N^2-1}{256 N^4}\Bigg\{ & & 
\left[
\frac{1261}{128} 
+\frac{50047}{384}  N^{2}
-\frac{66577}{1152}  N^{4}
+\frac{460151}{1152}  N^{6}
+21  \,\zeta_{3}
  \right. \nonumber \\ &{}& \left.  
\phantom{+}
-\frac{47}{2}  N^{2} \,\zeta_{3}
+52  N^{4} \,\zeta_{3}
+\frac{1157}{18}  N^{6} \,\zeta_{3}
-110  N^{4} \,\zeta_{5}
  \right. \nonumber \\ &{}& \left.  
\phantom{+}
-110  N^{6} \,\zeta_{5}
\right]
\nonumber\\
 &{+}&  \,n_f 
\left[
\frac{37}{6}  N 
+\frac{10475}{216}  N^{3}
-\frac{11908}{81}  N^{5}
-\frac{111}{2}  N  \,\zeta_{3}
  \right. \nonumber \\ &{}& \left.  
\phantom{+ \,n_f }
-85  N^{3} \,\zeta_{3}
-\frac{889}{6}  N^{5} \,\zeta_{3}
+33  N^{3} \,\zeta_{4}
+33  N^{5} \,\zeta_{4}
  \right. \nonumber \\ &{}& \left.  
\phantom{+ \,n_f }
-30  N  \,\zeta_{5}
+50  N^{3} \,\zeta_{5}
+80  N^{5} \,\zeta_{5}
\right]
\nonumber\\
 &{+}&  \, n_f^{2}
\left[
-\frac{19}{27}  N^{2}
+\frac{899}{324}  N^{4}
+10  N^{2} \,\zeta_{3}
+10  N^{4} \,\zeta_{3}
-6  N^{2} \,\zeta_{4}
  \right. \nonumber \\ &{}& \left.  
\phantom{+ \, n_f^{2}}
-6  N^{4} \,\zeta_{4}
\right]
\nonumber\\
 &{+}&  \, n_f^{3}
\left[
-\frac{83}{162}  N^{3}
+\frac{8}{9}  N^{3} \,\zeta_{3}
\right]
\,\Bigg\}{},
\label{MassAnomalDimMS3}
\end{eqnarray}

\begin{eqnarray}\gamma_{m}^{ \mathrm{RI}  (0)} = \frac{N^2-1}{2 N}\Bigg\{ & & 
 \frac{3}{4}\,\Bigg\}{},
\label{MassAnomalDimRI0}
\end{eqnarray}

\begin{eqnarray}\gamma_{m}^{ \mathrm{RI}  (1)} = \frac{N^2-1}{16 N^2}\Bigg\{ & & 
\left[
-\frac{3}{8} 
+\frac{379}{24}  N^{2}
\right]
 \, + \,  \,n_f 
\left[
-\frac{13}{6}  N 
\right]
\,\Bigg\}{},
\label{MassAnomalDimRI1}
\end{eqnarray}

\begin{eqnarray}\gamma_{m}^{ \mathrm{RI}  (2)} = \frac{N^2-1}{64 N^3}\Bigg\{ & & 
\left[
\frac{129}{16} 
-17  N^{2}
-22  \,\zeta_{3} N^{2}
+\frac{126239}{432}  N^{4}
-44  \,\zeta_{3} N^{4}
\right]
\nonumber\\
 &{+}&  \,n_f 
\left[
\frac{75}{8}  N 
-2  \,\zeta_{3} N 
-\frac{18611}{216}  N^{3}
+2  \,\zeta_{3} N^{3}
\right]
\nonumber\\
 &{+}&  \, n_f^{2}
\left[
\frac{116}{27}  N^{2}
\right]
\,\Bigg\}{},
\label{MassAnomalDimRI2}
\end{eqnarray}

\begin{eqnarray}\gamma_{m}^{ \mathrm{RI}  (3)} = \frac{N^2-1}{256 N^4}\Bigg\{ & & 
\left[
\frac{1261}{128} 
+\frac{198679}{384}  N^{2}
-\frac{778421}{1152}  N^{4}
+\frac{202580155}{31104}  N^{6}
  \right. \nonumber \\ &{}& \left.  
\phantom{+}
+21  \,\zeta_{3}
-\frac{149}{4}  N^{2} \,\zeta_{3}
-\frac{4133}{6}  N^{4} \,\zeta_{3}
-\frac{59269}{32}  N^{6} \,\zeta_{3}
  \right. \nonumber \\ &{}& \left.  
\phantom{+}
-165  N^{2} \,\zeta_{5}
+275  N^{4} \,\zeta_{5}
\right]
\nonumber\\
 &{+}&  \,n_f 
\left[
-\frac{3175}{48}  N 
+\frac{206575}{432}  N^{3}
-\frac{7671073}{2592}  N^{5}
  \right. \nonumber \\ &{}& \left.  
\phantom{+ \,n_f }
-59  N  \,\zeta_{3}
-\frac{22}{3}  N^{3} \,\zeta_{3}
+\frac{7767}{16}  N^{5} \,\zeta_{3}
  \right. \nonumber \\ &{}& \left.  
\phantom{+ \,n_f }
-20  N^{3} \,\zeta_{5}
+60  N^{5} \,\zeta_{5}
\right]
\nonumber\\
 &{+}&  \, n_f^{2}
\left[
-\frac{5767}{108}  N^{2}
+\frac{113747}{324}  N^{4}
+\frac{62}{3}  N^{2} \,\zeta_{3}
-32  N^{4} \,\zeta_{3}
\right]
\nonumber\\
 &{+}&  \, n_f^{3}
\left[
-\frac{2354}{243}  N^{3}
\right]
\,\Bigg\}{},
\label{MassAnomalDimRI3}
\end{eqnarray}

\begin{eqnarray}\gamma_{m}^{ \mathrm{RI'}  (0)} = \frac{N^2-1}{2 N}\Bigg\{ & & 
 \frac{3}{4}\,\Bigg\}{},
\label{MassAnomalDimRIp0}
\end{eqnarray}

\begin{eqnarray}\gamma_{m}^{ \mathrm{RI'}  (1)} = \frac{N^2-1}{16 N^2}\Bigg\{ & & 
\left[
-\frac{3}{8} 
+\frac{379}{24}  N^{2}
\right]
 \, + \,  \,n_f 
\left[
-\frac{13}{6}  N 
\right]
\,\Bigg\}{},
\label{MassAnomalDimRIp1}
\end{eqnarray}

\begin{eqnarray}\gamma_{m}^{ \mathrm{RI'}  (2)} = \frac{N^2-1}{64 N^3}\Bigg\{ & & 
\left[
\frac{129}{16} 
-\frac{147}{8}  N^{2}
-22  \,\zeta_{3} N^{2}
+\frac{121883}{432}  N^{4}
-44  \,\zeta_{3} N^{4}
\right]
\nonumber\\
 &{+}&  \,n_f 
\left[
\frac{77}{8}  N 
-2  \,\zeta_{3} N 
-\frac{17819}{216}  N^{3}
+2  \,\zeta_{3} N^{3}
\right]
\nonumber\\
 &{+}&  \, n_f^{2}
\left[
\frac{107}{27}  N^{2}
\right]
\,\Bigg\}{},
\label{MassAnomalDimRIp2}
\end{eqnarray}

\begin{eqnarray}\gamma_{m}^{ \mathrm{RI'}  (3)} = \frac{N^2-1}{256 N^4}\Bigg\{ & & 
\left[
\frac{1261}{128} 
+21  \,\zeta_{3}
+\frac{198283}{384}  N^{2}
-\frac{149}{4}  \,\zeta_{3} N^{2}
-165  \,\zeta_{5} N^{2}
  \right. \nonumber \\ &{}& \left.  
\phantom{+}
-\frac{844829}{1152}  N^{4}
-\frac{2017}{3}  \,\zeta_{3} N^{4}
+275  \,\zeta_{5} N^{4}
  \right. \nonumber \\ &{}& \left.  
\phantom{+}
+\frac{191436121}{31104}  N^{6}
-\frac{28551}{16}  \,\zeta_{3} N^{6}
\right]
\nonumber\\
 &{+}&  \,n_f 
\left[
-\frac{199}{3}  N 
-59  \,\zeta_{3} N 
+\frac{211669}{432}  N^{3}
-\frac{31}{3}  \,\zeta_{3} N^{3}
  \right. \nonumber \\ &{}& \left.  
\phantom{+ \,n_f }
-20  \,\zeta_{5} N^{3}
-\frac{1794277}{648}  N^{5}
+\frac{3697}{8}  \,\zeta_{3} N^{5}
  \right. \nonumber \\ &{}& \left.  
\phantom{+ \,n_f }
+60  \,\zeta_{5} N^{5}
\right]
\nonumber\\
 &{+}&  \, n_f^{2}
\left[
-\frac{1444}{27}  N^{2}
+\frac{62}{3}  \,\zeta_{3} N^{2}
+\frac{25946}{81}  N^{4}
-30  \,\zeta_{3} N^{4}
\right]
\nonumber\\
 &{+}&  \, n_f^{3}
\left[
-\frac{2003}{243}  N^{3}
\right]
\,\Bigg\}{},
\label{MassAnomalDimRIp3}
\end{eqnarray}

\begin{eqnarray}\gamma_{2}^{(0)} = & & \frac{N^2-1}{N}
\left[
\frac{1}{8}  \,\xi_{L} 
\right]
{},
\label{FieldAnomalDimMS0}
\end{eqnarray}

\begin{eqnarray}\gamma_{2}^{(1)} = \frac{N^2-1}{16 N^2}\Bigg\{ & & 
\left[
\frac{3}{8} 
+\frac{11}{4}  N^{2}
+  N^{2} \,\xi_{L} 
+\frac{1}{8}  N^{2} \xi_{L}^{2}
\right]
 \, + \,  \,n_f 
\left[
-\frac{1}{2}  N 
\right]
\,\Bigg\}{},
\label{FieldAnomalDimMS1}
\end{eqnarray}

\begin{eqnarray}\gamma_{2}^{(2)} = \frac{N^2 - 1}{64 N^3}\Bigg\{ & & 
\left[
\frac{3}{16} 
+\frac{137}{16}  N^{2}
+\frac{6635}{288}  N^{4}
+\frac{263}{64}  N^{4} \,\xi_{L} 
+\frac{39}{64}  N^{4} \xi_{L}^{2}
  \right. \nonumber \\ &{}& \left.  
\phantom{+}
+\frac{5}{32}  N^{4} \xi_{L}^{3}
-3  N^{2} \,\zeta_{3}
-\frac{21}{16}  N^{4} \,\zeta_{3}
+\frac{3}{8}  N^{4} \,\xi_{L}  \,\zeta_{3}
  \right. \nonumber \\ &{}& \left.  
\phantom{+}
+\frac{3}{16}  N^{4} \xi_{L}^{2} \,\zeta_{3}
\right]
\nonumber\\
 &{+}&  \,n_f 
\left[
-\frac{3}{8}  N 
-\frac{547}{72}  N^{3}
-\frac{17}{16}  N^{3} \,\xi_{L} 
\right]
\nonumber\\
 &{+}&  \, n_f^{2}
\left[
\frac{5}{18}  N^{2}
\right]
\,\Bigg\}{},
\label{FieldAnomalDimMS2}
\end{eqnarray}

\begin{eqnarray}\gamma_{2}^{(3)} = \frac{N^2 - 1}{256 N^4}\Bigg\{ & & 
\left[
\frac{1027}{128} 
+\frac{11281}{384}  N^{2}
+\frac{86017}{1152}  N^{4}
+\frac{1785121}{10368}  N^{6}
  \right. \nonumber \\ &{}& \left.  
\phantom{+}
-\frac{5}{8}  N^{4} \,\xi_{L} 
+\frac{1644899}{62208}  N^{6} \,\xi_{L} 
+\frac{6467}{2304}  N^{6} \xi_{L}^{2}
  \right. \nonumber \\ &{}& \left.  
\phantom{+}
+\frac{149}{192}  N^{6} \xi_{L}^{3}
+\frac{19}{128}  N^{6} \xi_{L}^{4}
+25  \,\zeta_{3}
+31  N^{2} \,\zeta_{3}
  \right. \nonumber \\ &{}& \left.  
\phantom{+}
-\frac{4287}{64}  N^{4} \,\zeta_{3}
-\frac{1103}{64}  N^{6} \,\zeta_{3}
+\frac{115}{16}  N^{4} \,\xi_{L}  \,\zeta_{3}
  \right. \nonumber \\ &{}& \left.  
\phantom{+}
+\frac{2761}{192}  N^{6} \,\xi_{L}  \,\zeta_{3}
+\frac{11}{32}  N^{4} \xi_{L}^{2} \,\zeta_{3}
+\frac{289}{96}  N^{6} \xi_{L}^{2} \,\zeta_{3}
  \right. \nonumber \\ &{}& \left.  
\phantom{+}
-\frac{3}{16}  N^{4} \xi_{L}^{3} \,\zeta_{3}
+\frac{19}{64}  N^{6} \xi_{L}^{3} \,\zeta_{3}
-\frac{7}{64}  N^{4} \xi_{L}^{4} \,\zeta_{3}
  \right. \nonumber \\ &{}& \left.  
\phantom{+}
-\frac{1}{64}  N^{6} \xi_{L}^{4} \,\zeta_{3}
+\frac{33}{2}  N^{4} \,\zeta_{4}
+\frac{231}{32}  N^{6} \,\zeta_{4}
-\frac{97}{64}  N^{6} \,\xi_{L}  \,\zeta_{4}
  \right. \nonumber \\ &{}& \left.  
\phantom{+}
-\frac{17}{32}  N^{6} \xi_{L}^{2} \,\zeta_{4}
-\frac{3}{64}  N^{6} \xi_{L}^{3} \,\zeta_{4}
-40  \,\zeta_{5}
-60  N^{2} \,\zeta_{5}
  \right. \nonumber \\ &{}& \left.  
\phantom{+}
+\frac{1945}{64}  N^{4} \,\zeta_{5}
-\frac{1375}{128}  N^{6} \,\zeta_{5}
-\frac{65}{8}  N^{4} \,\xi_{L}  \,\zeta_{5}
  \right. \nonumber \\ &{}& \left.  
\phantom{+}
-\frac{95}{8}  N^{6} \,\xi_{L}  \,\zeta_{5}
-\frac{35}{32}  N^{4} \xi_{L}^{2} \,\zeta_{5}
-\frac{75}{64}  N^{6} \xi_{L}^{2} \,\zeta_{5}
  \right. \nonumber \\ &{}& \left.  
\phantom{+}
+\frac{5}{64}  N^{4} \xi_{L}^{4} \,\zeta_{5}
+\frac{5}{128}  N^{6} \xi_{L}^{4} \,\zeta_{5}
\right]
\nonumber\\
 &{+}&  \,n_f 
\left[
\frac{307}{12}  N 
-\frac{1555}{144}  N^{3}
-\frac{35641}{432}  N^{5}
+\frac{767}{96}  N^{3} \,\xi_{L} 
  \right. \nonumber \\ &{}& \left.  
\phantom{+ \,n_f }
-\frac{161347}{15552}  N^{5} \,\xi_{L} 
-\frac{109}{288}  N^{5} \xi_{L}^{2}
-4  N  \,\zeta_{3}
+8  N^{3} \,\zeta_{3}
  \right. \nonumber \\ &{}& \left.  
\phantom{+ \,n_f }
-\frac{35}{8}  N^{5} \,\zeta_{3}
-\frac{11}{2}  N^{3} \,\xi_{L}  \,\zeta_{3}
-\frac{15}{4}  N^{5} \,\xi_{L}  \,\zeta_{3}
  \right. \nonumber \\ &{}& \left.  
\phantom{+ \,n_f }
-\frac{7}{24}  N^{5} \xi_{L}^{2} \,\zeta_{3}
-3  N^{3} \,\zeta_{4}
-\frac{21}{16}  N^{5} \,\zeta_{4}
-\frac{3}{2}  N^{3} \,\xi_{L}  \,\zeta_{4}
  \right. \nonumber \\ &{}& \left.  
\phantom{+ \,n_f }
-\frac{1}{2}  N^{5} \,\xi_{L}  \,\zeta_{4}
+\frac{1}{16}  N^{5} \xi_{L}^{2} \,\zeta_{4}
-20  N^{3} \,\zeta_{5}
\right]
\nonumber\\
 &{+}&  \, n_f^{2}
\left[
-\frac{19}{9}  N^{2}
+\frac{445}{72}  N^{4}
-\frac{269}{486}  N^{4} \,\xi_{L} 
+2  N^{2} \,\zeta_{3}
+2  N^{4} \,\zeta_{3}
  \right. \nonumber \\ &{}& \left.  
\phantom{+ \, n_f^{2}}
+\frac{2}{3}  N^{4} \,\xi_{L}  \,\zeta_{3}
\right]
 \, + \,  \, n_f^{3}
\left[
\frac{35}{162}  N^{3}
\right]
\,\Bigg\}{},
\label{FieldAnomalDimMS3}
\end{eqnarray}

\begin{eqnarray}\gamma_{2}^{ \mathrm{RI}  (1)} = \frac{N^2-1}{16 N^2}\Bigg\{ & & 
\left[
\frac{3}{8} 
+\frac{11}{4}  N^{2}
\right]
 \, + \,  \,n_f 
\left[
-\frac{1}{2}  N 
\right]
\,\Bigg\}{},
\label{FieldAnomalDimRI1}
\end{eqnarray}

\begin{eqnarray}\gamma_{2}^{ \mathrm{RI}  (2)} = \frac{N^2 - 1}{64 N^3}\Bigg\{ & & 
\left[
\frac{3}{16} 
+\frac{25}{3}  N^{2}
+\frac{14225}{288}  N^{4}
-3  N^{2} \,\zeta_{3}
-\frac{197}{16}  N^{4} \,\zeta_{3}
\right]
\nonumber\\
 &{+}&  \,n_f 
\left[
-\frac{1}{3}  N 
-\frac{611}{36}  N^{3}
+2  N^{3} \,\zeta_{3}
\right]
\nonumber\\
 &{+}&  \, n_f^{2}
\left[
\frac{10}{9}  N^{2}
\right]
\,\Bigg\}{},
\label{FieldAnomalDimRI2}
\end{eqnarray}

\begin{eqnarray}\gamma_{2}^{ \mathrm{RI}  (3)} = \frac{N^2 - 1}{256 N^4}\Bigg\{ & & 
\left[
\frac{1027}{128} 
+\frac{7673}{384}  N^{2}
+\frac{174565}{1152}  N^{4}
+\frac{3993865}{3456}  N^{6}
  \right. \nonumber \\ &{}& \left.  
\phantom{+}
+25  \,\zeta_{3}
+31  N^{2} \,\zeta_{3}
-\frac{10975}{64}  N^{4} \,\zeta_{3}
-\frac{111719}{192}  N^{6} \,\zeta_{3}
  \right. \nonumber \\ &{}& \left.  
\phantom{+}
-40  \,\zeta_{5}
-60  N^{2} \,\zeta_{5}
+\frac{5465}{64}  N^{4} \,\zeta_{5}
+\frac{20625}{128}  N^{6} \,\zeta_{5}
\right]
\nonumber\\
 &{+}&  \,n_f 
\left[
\frac{1307}{48}  N 
+\frac{557}{144}  N^{3}
-\frac{172793}{288}  N^{5}
  \right. \nonumber \\ &{}& \left.  
\phantom{+ \,n_f }
-4  N  \,\zeta_{3}
+2  N^{3} \,\zeta_{3}
+\frac{7861}{48}  N^{5} \,\zeta_{3}
  \right. \nonumber \\ &{}& \left.  
\phantom{+ \,n_f }
-30  N^{3} \,\zeta_{5}
-\frac{125}{4}  N^{5} \,\zeta_{5}
\right]
\nonumber\\
 &{+}&  \, n_f^{2}
\left[
-\frac{521}{72}  N^{2}
+\frac{259}{3}  N^{4}
+6  N^{2} \,\zeta_{3}
-\frac{26}{3}  N^{4} \,\zeta_{3}
\right]
\nonumber\\
 &{+}&  \, n_f^{3}
\left[
-\frac{86}{27}  N^{3}
\right]
\,\Bigg\}{},
\label{FieldAnomalDimRI3}
\end{eqnarray}

\begin{eqnarray}\gamma_{2}^{ \mathrm{RI'}  (1)} = \frac{N^2-1}{16 N^2}\Bigg\{ & & 
\left[
\frac{3}{8} 
+\frac{11}{4}  N^{2}
\right]
 \, + \,  \,n_f 
\left[
-\frac{1}{2}  N 
\right]
\,\Bigg\}{},
\label{FieldAnomalDimRIp1}
\end{eqnarray}

\begin{eqnarray}\gamma_{2}^{ \mathrm{RI'}  (2)} = \frac{N^2 - 1}{64 N^3}\Bigg\{ & & 
\left[
\frac{3}{16} 
+\frac{233}{24}  N^{2}
+\frac{17129}{288}  N^{4}
-3  N^{2} \,\zeta_{3}
-\frac{197}{16}  N^{4} \,\zeta_{3}
\right]
\nonumber\\
 &{+}&  \,n_f 
\left[
-\frac{7}{12}  N 
-\frac{743}{36}  N^{3}
+2  N^{3} \,\zeta_{3}
\right]
\nonumber\\
 &{+}&  \, n_f^{2}
\left[
\frac{13}{9}  N^{2}
\right]
\,\Bigg\}{},
\label{FieldAnomalDimRIp2}
\end{eqnarray}

\begin{eqnarray}\gamma_{2}^{ \mathrm{RI'}  (3)} = \frac{N^2 - 1}{256 N^4}\Bigg\{ & & 
\left[
\frac{1027}{128} 
+\frac{8069}{384}  N^{2}
+\frac{240973}{1152}  N^{4}
+\frac{5232091}{3456}  N^{6}
  \right. \nonumber \\ &{}& \left.  
\phantom{+}
+25  \,\zeta_{3}
+31  N^{2} \,\zeta_{3}
-\frac{12031}{64}  N^{4} \,\zeta_{3}
-\frac{124721}{192}  N^{6} \,\zeta_{3}
  \right. \nonumber \\ &{}& \left.  
\phantom{+}
-40  \,\zeta_{5}
-60  N^{2} \,\zeta_{5}
+\frac{5465}{64}  N^{4} \,\zeta_{5}
+\frac{20625}{128}  N^{6} \,\zeta_{5}
\right]
\nonumber\\
 &{+}&  \,n_f 
\left[
\frac{329}{12}  N 
-\frac{1141}{144}  N^{3}
-\frac{113839}{144}  N^{5}
  \right. \nonumber \\ &{}& \left.  
\phantom{+ \,n_f }
-4  N  \,\zeta_{3}
+5  N^{3} \,\zeta_{3}
+\frac{2245}{12}  N^{5} \,\zeta_{3}
  \right. \nonumber \\ &{}& \left.  
\phantom{+ \,n_f }
-30  N^{3} \,\zeta_{5}
-\frac{125}{4}  N^{5} \,\zeta_{5}
\right]
\nonumber\\
 &{+}&  \, n_f^{2}
\left[
-\frac{515}{72}  N^{2}
+\frac{1405}{12}  N^{4}
+6  N^{2} \,\zeta_{3}
-\frac{32}{3}  N^{4} \,\zeta_{3}
\right]
\nonumber\\
 &{+}&  \, n_f^{3}
\left[
-\frac{125}{27}  N^{3}
\right]
\,\Bigg\}{}.
\label{FieldAnomalDimRIp3}
\end{eqnarray}

\end{appendix}

\newpage

\def\app#1#2#3{{\it Act.~Phys.~Pol.~}{\bf B #1} (#2) #3}
\def\apa#1#2#3{{\it Act.~Phys.~Austr.~}{\bf#1} (#2) #3}
\def\cmp#1#2#3{{\it Comm.~Math.~Phys.~}{\bf #1} (#2) #3}
\def\cpc#1#2#3{{\it Comp.~Phys.~Commun.~}{\bf #1} (#2) #3}
\def\epjc#1#2#3{{\it Eur.\ Phys.\ J.\ }{\bf C #1} (#2) #3}
\def\fortp#1#2#3{{\it Fortschr.~Phys.~}{\bf#1} (#2) #3}
\def\ijmpa#1#2#3{{\it Int.~J.~Mod.~Phys.~}{\bf A #1} (#2) #3}
\def\jcp#1#2#3{{\it J.~Comp.~Phys.~}{\bf #1} (#2) #3}
\def\jetp#1#2#3{{\it JETP~Lett.~}{\bf #1} (#2) #3}
\def\mpl#1#2#3{{\it Mod.~Phys.~Lett.~}{\bf A #1} (#2) #3}
\def\nima#1#2#3{{\it Nucl.~Inst.~Meth.~}{\bf A #1} (#2) #3}
\def\npb#1#2#3{{\it Nucl.~Phys.~}{\bf B #1} (#2) #3}
\def\nca#1#2#3{{\it Nuovo~Cim.~}{\bf #1A} (#2) #3}
\def\plb#1#2#3{{\it Phys.~Lett.~}{\bf B #1} (#2) #3}
\def\prc#1#2#3{{\it Phys.~Reports }{\bf #1} (#2) #3}
\def\prd#1#2#3{{\it Phys.~Rev.~}{\bf D #1} (#2) #3}
\def\pR#1#2#3{{\it Phys.~Rev.~}{\bf #1} (#2) #3}
\def\prl#1#2#3{{\it Phys.~Rev.~Lett.~}{\bf #1} (#2) #3}
\def\pr#1#2#3{{\it Phys.~Reports }{\bf #1} (#2) #3}
\def\ptp#1#2#3{{\it Prog.~Theor.~Phys.~}{\bf #1} (#2) #3}
\def\sovnp#1#2#3{{\it Sov.~J.~Nucl.~Phys.~}{\bf #1} (#2) #3}
\def\tmf#1#2#3{{\it Teor.~Mat.~Fiz.~}{\bf #1} (#2) #3}
\def\tmph#1#2#3{{\it Theor.~Math.~Phys.}{\bf #1} (#2) #3}
\def\yadfiz#1#2#3{{\it Yad.~Fiz.~}{\bf #1} (#2) #3}
\def\zpc#1#2#3{{\it Z.~Phys.~}{\bf C #1} (#2) #3}
\def\ibid#1#2#3{{ibid.~}{\bf #1} (#2) #3}

\end{document}